\documentclass[review]{elsarticle}
\usepackage{amsmath}
\usepackage{amsfonts}
\usepackage{amssymb}
\usepackage[english]{babel}
\usepackage{graphicx}
\usepackage{color} 
\usepackage{todonotes}
\usepackage{cleveref}
\usepackage{array}
\usepackage{verbatim}

\usepackage[caption=false]{subfig}

\newcommand{\code}{$\mathtt{FREYA}$}
\newcommand{\sci}{\text{sc}}
\newcommand{\stat}{\text{stat}}
\newcommand{\rot}{\text{rot}}
\newcommand{\TKE}{\text{TKE}}
\newcommand{\cf}{$^{252}$Cf(sf)}
\newcommand{\ur}{$^{238}$U(sf)}
\newcommand{\cm}{$^{244}$Cm(sf)}
\newcommand{\pu}[1]{$^{#1}$Pu(sf)}
\newcommand{\lr}[1]{\left\langle #1 \right\rangle}
\newcommand{\mev}{\ \textrm{MeV}}
\newcommand{\imev}{/\textrm{MeV}}
\graphicspath{{visuals/}}
\newcolumntype{L}{>{$}l<{$}}
\newcolumntype{C}{>{$}c<{$}}
\newcolumntype{R}{>{$}r<{$}}
\Crefname{section}{Sec.}{Sections}
\Crefname{figure}{Fig.}{Figs.}
\Crefname{eqn}{Eq.}{Eqs.}

\begin{document}

\title{Parameter Optimization and Uncertainty Analysis of \code\ for Spontaneous Fission}

\author[1]{J. Van Dyke}
\author[2,3]{L. A. Bernstein}
\author[4,5]{R. Vogt}

\address[1]{Physics Department, University of California, Berkeley, CA 94720}
\address[2]{Nuclear Science Division, Lawrence Berkeley National Laboratory, Berkeley, CA 94720}
\address[3]{Nuclear Engineering Department, University of California, Berkeley, CA 95616}
\address[4]{Nuclear and Chemical Sciences Division, Lawrence Livermore National Laboratory, Livermore CA 94551}
\address[5]{Physics Department, University of California, Davis, CA 95616}

\date{\today}

\begin{abstract}
In this paper we report on an effort to determine an optimal parameter
set for the complete event fission model \code\ 
to reproduce spontaneous fission of \cf, \cm, \pu{238}, \pu{240}, \pu{242},
and \ur.
Earlier studies have partially optimized the event-by-event fission model
\code\ with respect to the available experimental data using brute force 
computational techniques. 
We have confirmed and expanded these results using a least-squares
minimization based on the simulated annealing approach. 
We have also developed a more complete statistical picture of this optimization, 
consisting of a full correlation matrix for the parameters utilized by \code.
The newly improved parameter values themselves, along with this correlation matrix,
have led to a more well-developed physical picture of the fission process.
\end{abstract}

\maketitle

\section{Introduction}
\label{sec:intro}

Though nuclear fission has influenced society in significant ways,
the fission process itself is still not understood in great detail.
Nevertheless, we can produce a complete fully-correlated
physically-consistent description of fission.
% to aid in the design and understanding of systems where it plays a significant role.
The Fission Reaction Event Yield Algorithm (\code) fission model
is designed to serve this purpose in a physically-complete fashion with
a relatively modest computational footprint.
While our main focus here is on \cf,
we also present optimized parameters for all spontaneously
fissioning nuclei in the event-by-event simulation code
\code\ \cite{VR_O,RVJR_spont,VRBDO,RVJR_gamma,CPC,CPC_NVA}.
\code\ generates samples of complete fission events,
including the full kinematic information for the two product nuclei,
as well as the emitted neutrons and photons.
It was designed to quickly generate large numbers of events.
\code\ is also a published code \cite{CPC,CPC_NVA},
In this work we concern ourselves with improving the \code\ input parameters
by global optimization and considering the
physical implications of the resulting parameter values.
We also discuss the areas in which experimental data is lacking
for this type of comparison.

The events generated by \code\ depend on five physics-based parameters.
For a given choice of these five parameters, the results can be compared to
existing experimental data and evaluations in order to determine how effective
each choice is at describing all the data.
This work was carried out using multiple different numerical optimization techniques
in order to determine the most efficient and effective methodology.
We make a full statistical analysis, including variances and covariances.
In this paper we find the best possible set of the input parameters
for describing all the spontaneous fission data for each isotope.

In \Cref{sec:parameters} we describe the parameters we optimize in \code.
\Cref{sec:computational} discusses the numerical methods used to perform the optimization,
while \Cref{sec:data} identifies the data employed in the fits.
\Cref{sec:results,sec:interpretation} provide the results and their interpretation,
as well as a comparison between the resulting parameter values to those previously used.
We compare results to the data for specific \cf\ observables in \Cref{sec:comp_to_data}.
Comparisons to other isotopes can be found in the Appendix.

\section{\code\ parameter description}
\label{sec:parameters}

In this section we briefly discuss the process of nuclear fission as implemented in \code.
We also identify and provide a physical interpretation of
the five parameters required by \code.

The fission process begins when a specified initial compound nucleus
splits into two fragment nuclei, typically one
light and one heavy, which we denote by $L$ and $H$
respectively for each fragment pair.
The corresponding $Q$-value is given by $Q = M_0 c^2 - M_L c^2 - M_H c^2$
for spontaneous fission.
The fragment yields as a function of fragment mass and the total kinetic
energy of the fragments, $\TKE$,
as a function of heavy fragment mass, $A_H$, are sampled from data.
From the fission $Q$ value and the sampled $\TKE$ we determine the total
excitation energy at scission, $E_{\sci}^*$, by energy conservation.
The excitation energy $E_{\sci}^*$ is available for both statistical,
$E_{\stat}$, and rotational, $E_{\rot}$, excitation of the fragments.
These two quantities are related by:
\begin{equation}
E_{\sci}^* = Q - \overline{\TKE} =
E_\stat + E_\rot \ .
\end{equation}
The level density parameter\footnote{
The relation given here is an approximation valid for high energies and negligible
shell corrections. In \code\ a back-shifted Fermi gas model is used. See
Ref.~\cite{VRBDO} for details.}
$a \approx A_0 / e_0$ \ \cite{RVJR_spont},
for some constant $e_0$,
determines a ``scission temperature'' $T_{\sci}$ from the relation:
\begin{equation}
E_{\sci}^* = a T_{\sci}^2 \ .
\label{Eq:scission}
\end{equation}
This $e_0$ is the first parameter required by \code, and is usually around
$10 \imev$
\cite{CPC_NVA}.  Note that, while Eq.~(\ref{Eq:scission}) relates $a$ to the
scission temperature, the level density parameter $a$ is also employed for all
neutron emission during the fission process.

In addition to the mean angular momenta of the fragment given by the overall
rigid rotation around the scission axis,
there are also fluctuations around this value attributed to the wriggling and
bending modes \cite{JRRV_ang} that contribute to $E_{\rot}$.
The relative degree of these fluctuations is given by
\begin{equation}
T_S = c_S T_{\sci} \ .
\label{eqn:c_S}
\end{equation}
The ratio of the fluctuation temperature $T_S$
to the scission temperature $T_{\sci}$, $c_S$, is our second parameter.
It is clear that this must be non-zero.
If it were zero, there would be no fluctuations and the only angular momentum
present in the fragments would be that dictated by the rigid rotation
before scission. In the case of spontaneous fission, this would
mean that the fragments have no angular momentum, which is not the case.
The default value used in the most recently published version of \code\ is $c_S= 0.87$
\cite{CPC_NVA,Andrew}. See \cite{JRRV_ang} for more details on the addition of
angular momentum to \code.

The statistical excitation energy, $E_\stat$, is initially partitioned as
$E_\stat =  \acute E_L^* + \acute E_H^*$ where the $*$ indicates
that the statistical excitation is initially partitioned according
to the level density parameters.
This would only be completely accurate if the fragments were in mutual thermal
equilibrium.  However, since we know that the light fragment emits more
neutrons on average, we modify the partition via the third parameter, $x$:
\begin{align}
\overline{E}_L^* = x \acute E_L^* \ ,
&&
\overline{E}^*_H = E_\stat - \overline{E}_L^* \ ,
\label{eqn:x}
\end{align}
assumed to be greater than $1$.
A value around $1.1-1.3$ is typically found
\cite{Andrew,jerome_angular}.

As noted in Eq.~\eqref{Eq:scission}, the average fragment excitation energy is
proportional to the temperature, i.e. $\overline{E_i}^* \propto T_i^2$.
The variance of this excitation is given by:
\begin{equation}
\sigma_{E_i}^2 = c\overline{E_i}^* T_i \ .
\label{eqn:c}
\end{equation}
Therefore we have an energy fluctuation, written $\delta E_i^*$,
on both the heavy and light fragments.
This fluctuation is sampled from a normal distribution of variance equal to
$2c \overline{E_i}^* T_i$.
In particular, the excitation energy of each fragment is adjusted to be
$E_i^* = \overline{ E_i}^* + \delta E_i^*$.
Therefore we can understand the factor $c$, our fourth parameter,
as controlling the truncation of the normal
distribution at the maximum available excitation. It primarily affects the
neutron multiplicity distribution and was assumed to take a value $c \sim 1$.
We maintain energy conservation by
\begin{equation}
\TKE = \overline{\TKE} - \delta E_L^* - \delta E_H^* \, \, .
\end{equation}
Finally, to ensure reproduction of the measured average neutron multiplicity
$\overline\nu$, we allow the value of the average total kinetic energy
to shift by a small amount $d\TKE$.
The measured data have often unquantified systematic uncertainties or, in some cases,
low statistics.

The ranges considered for these parameters can be found in
\Cref{table:parameter_ranges}.
While the range for $c$ is listed as $1-3$, for some isotopes
we allow this range to expand.
Since this parameter controls the width of the neutron multiplicity distribution, for isotopes
which are known to have a comparatively narrow distribution, we allow the parameter to vary below $1$
to $0.8$. In addition, for isotopes with a comparatively wide distribution relative to their
average multiplicity we allow $c$ to be as
large as $4$.

We note that there are two detector-based photon-related parameters in \code,
$g_{\text{min}}$, the minimum detected photon energy, and $t_{\text{max}}$,
the length of the time measurement. Because these are unique to each measurement, they are not
counted as tunable parameters.
They do however have some effect on the photon multiplicity and energy per photon
\cite{RVJR_newgamma}.
The fits use the values of $g_{\text{min}}$ and $t_{\text{max}}$ appropriate for the data
included in the fits.

\begin{table}
{\footnotesize
\begin{tabular}{|C|C|C|C|C|}
\hline
e_0\, (/{\rm MeV}) & x & c_S & c & d\TKE \, ({\rm MeV}) \\
\hline
\hline
7 \text{ - } 12 &
1.0 \text{ - } 1.5 &
0.5 \text{ - } 1.5&
1 \text{ - } 3 &
-5 \text{ - } 5
\\ \hline
\end{tabular}
}
\caption{Ranges of parameters considered in the optimization.}
\label{table:parameter_ranges}
\end{table}

\section{Computational methods}
\label{sec:computational}

\begin{figure}[t]
\centering
\includegraphics[width=0.49\textwidth]{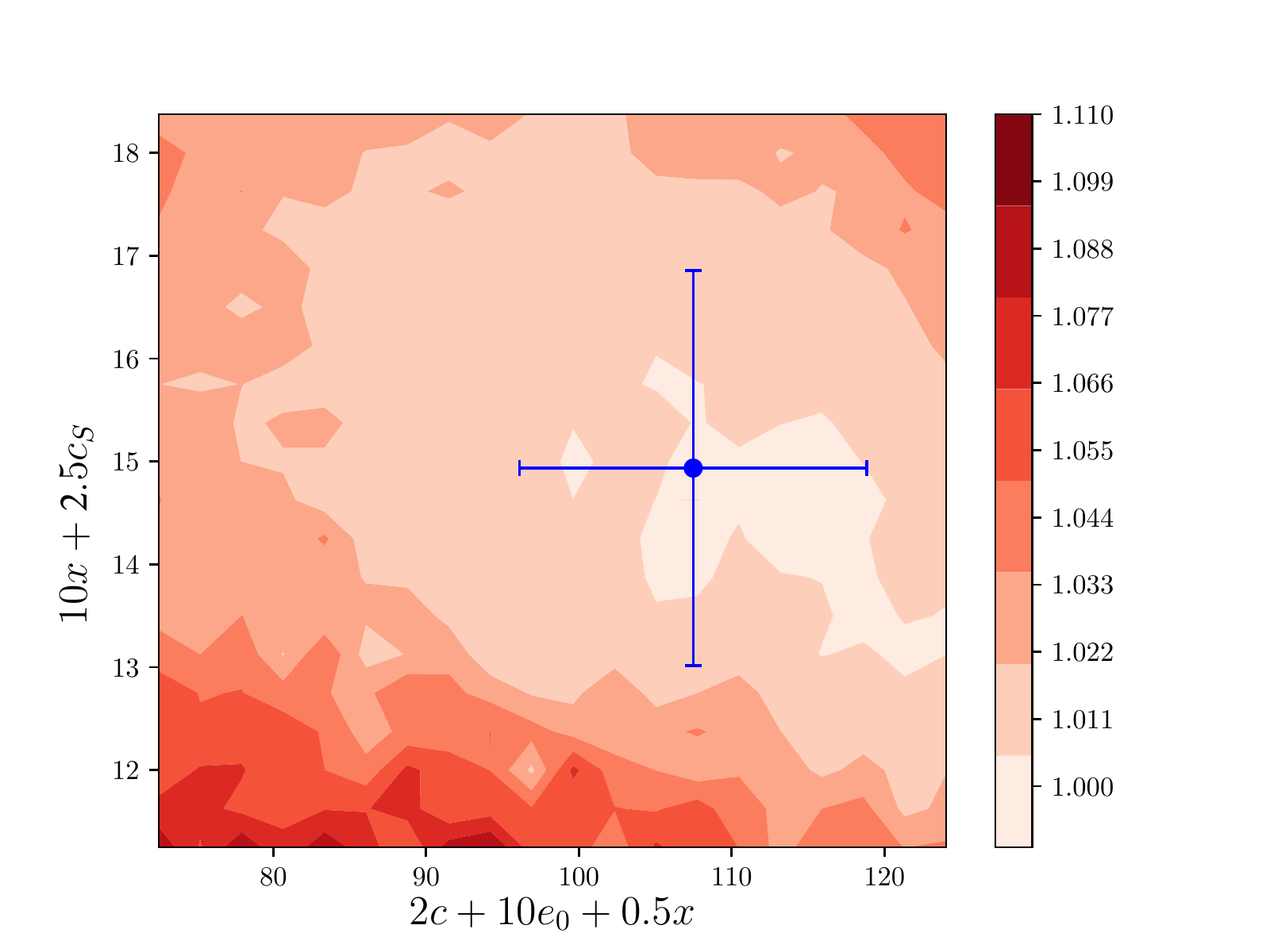}
\\
\includegraphics[width=0.49\textwidth]{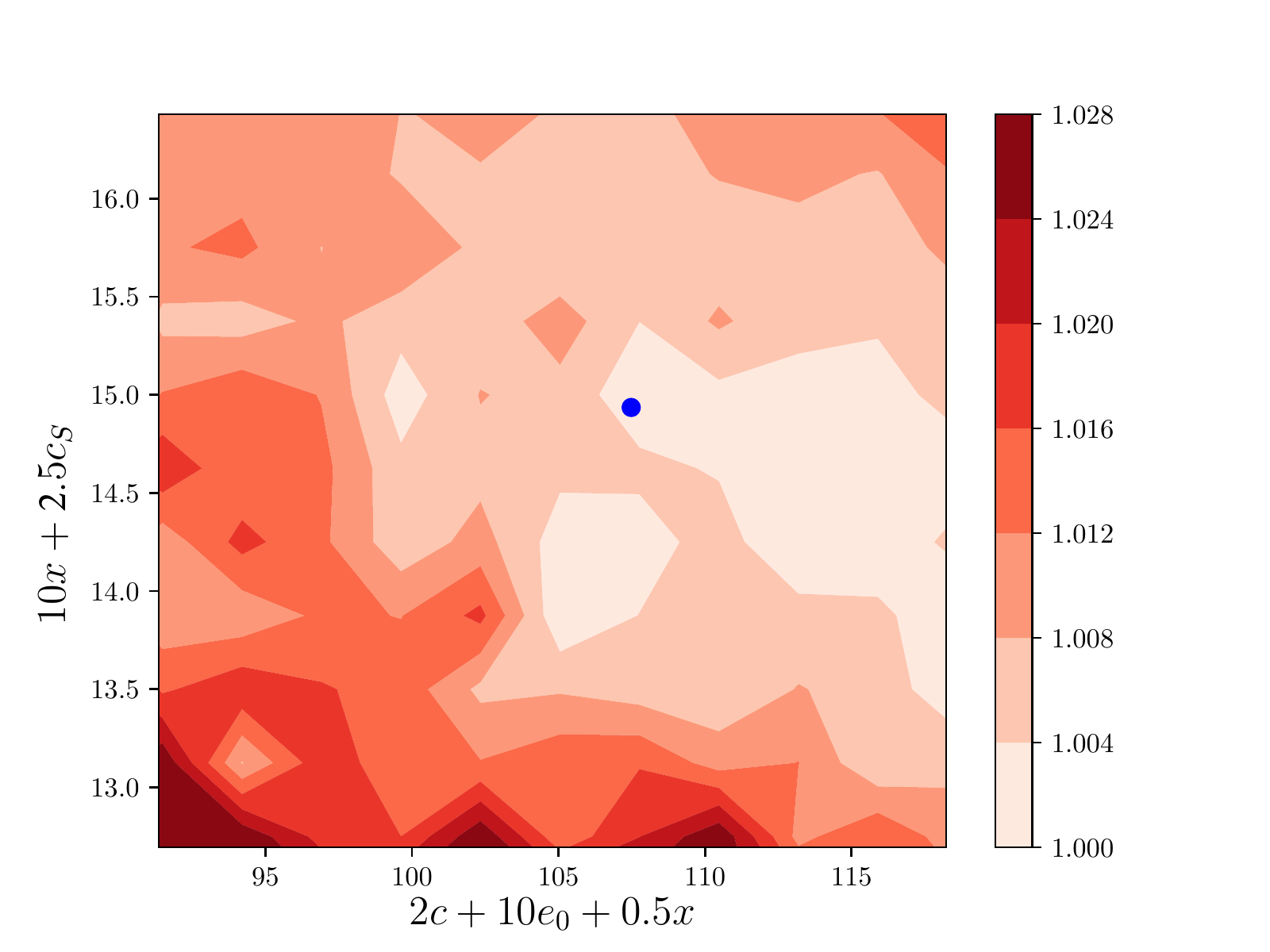}
\caption{(Color online) (a) Contour plot of $\chi^2$ relative to
various parameter values for \cf\ for the full range of tested parameters.
(b) Contour zoomed in to focus around optimized parameter values.}
\label{fig:contours_cf}
\end{figure}

For each set of five parameters, we generate sets of $1,000,000$ events.
The output from the generated events contains the full kinematic information for
the fragments and the emitted neutrons and photons.
We use this kinematic information to calculate
physical observable which are then compared to measured data.
In this study, the quantities we extracted included the average neutron
multiplicity, $\overline \nu$;
the second and third moments of the neutron multiplicity, $\nu_2$ and $\nu_3$
respectively; the neutron multiplicity distribution, $P(\nu)$;
the average neutron multiplicity as a function of the total kinetic energy,
$\nu({\rm TKE})$, and as a function of fragment mass, $\nu(A)$;
the neutron energy spectrum, $N(E)$; the average photon multiplicity,
$\overline N_\gamma$;
the photon multiplicity distribution, $P(N_\gamma)$;
and the average energy per photon $\overline \epsilon_\gamma$.

The moments of the multiplicity distribution are defined as
  \begin{equation}
    \nu_n = \sum_n \frac{\nu !}{(\nu - n) !} P(\nu)
  \end{equation}
  where
%In addition to the average neutron multiplicity, we also compare with the so-called second and third moments of the neutron multiplicity which are defined as:
\begin{eqnarray}
  \nu_1 & = & \overline \nu = \langle \nu \rangle \, \, , \nonumber \\
\nu_2 & = & \left\langle\nu\left( \nu-1 \right)\right\rangle \, \, , \nonumber \\
\nu_3 & = & \left\langle\nu\left( \nu-1 \right)\left( \nu - 2 \right)\right\rangle \, .
\label{eqn:moments}
\end{eqnarray}
These moments
numerically encapsulate the shape of the neutron multiplicity distribution.
After calculating these observables from the \code\ output, they are compared
with available experimental data and evaluations.
Unfortunately, not all of these observables are available for many of the isotopes of interest,
as we discuss later.
More information on the sources and quality of these data can be found in
\Cref{sec:data}.

The \code\ output is compared to the data, and for each observable
the reduced $\chi^2$, $\chi_0^2$ is calculated as
\begin{equation}
\chi^2_O = \frac{1}{n-5} \sum_{i = 1}^n
\frac{\left( O_i - E_i \right)^2}{\sigma_i^2}
\end{equation}
where $i = 1 , \dots , n$ runs over the bins of the distribution; $O_i$ is the
value of the observable
returned by \code\ for the given bin; $E_i$ is the experimental result;
and $\sigma_i$ is the experimental uncertainty on $E_i$.
The reduced $\chi_0^2$ for the observable is found by dividing the sum over all bins
by the number of degrees of freedom, $n-5$,
for the five physics-based parameters we are fitting.
For single valued observables such as $\overline \nu$, we simply take
$\chi_O^2 = \left( O - E \right)^2 / \sigma^2$.

The total $\chi^2$ is the sum over all observables where data are available,
\begin{equation}
\chi^2 = \sum_O \chi^2_O
\, \, . \label{totalchisq}
\end{equation}
This total $\chi^2$ is treated as the return value of an objective function.
In \Cref{fig:contours_cf} we plot the following:
first we take the sum of the reduced $\chi^2$ values for a linear combinations of the parameters.
Then we find the lowest such value, and plot the ratio of all of the values to this value.
These plots show us merely one particular two dimensional projection
of the five-dimensional parameter space.
The particular linear combination of parameters was
chosen by-eye to best illustrate the nature of the parameter
space we are working in.

Preliminary work in Ref.~\cite{Andrew} used a grid-search method where
every potential combination of parameters was tested.
However, in a five-dimensional space with a reasonably fine mesh, the grid search
technique is unwieldy and very computationally intensive. We have therefore
also used alternative methods and confirmed that our alternate methodology agreed
with the earlier used grid search approach used earlier.

\Cref{fig:contours_cf} shows that the objective function
displays many local minima which are neither global minima, nor physically relevant.
Therefore, we cannot employ a simple algorithm, such as gradient descent,
because it can easily fall into such local minima.
We have instead employed the so-called simulated annealing method \cite{anneal}.
The motivation for such an algorithm is to
inject a certain amount of randomness into the process to allow
for the procedure to occasionally jump in a seemingly ``worse'' direction
in order to move out of a potential local minimum and
eventually find the global solution.
We provide a rough description of the algorithm now.

The simulated annealing algorithm first generates a random solution,
calculates its cost using an objective function,
generates a random neighboring solution, calculates the cost of this new solution
with the same objective function, and then compares these costs using an
acceptance probability function.
The acceptance probability is calculated by comparing
the difference of the two costs with the so-called temperature, $T$.
The parameter $T$ is initially equal to unity,
and is decreased to a new value, $T'$,
after each iteration of the algorithm by employing a scale factor $\alpha$,
\begin{equation}
T' = \alpha T
\, \, . \label{temperature}
\end{equation}
The factor $\alpha$ is usually greater than $0.8$, and is always less than $1$.
The temperature allows for the algorithm to become less stochastic as the
number of iterations is increased.
The value returned by the acceptance probability function
is then compared to a randomly generated number to determine whether the
new solution is accepted.
As a result, when the algorithm compares the costs of these two solutions,
there is a certain probability that, even if the new solution is worse,
it still might be accepted.
This helps prevent the algorithm from sinking into a local minimum.
The process is repeated until an acceptable solution is found.

In our particular situation, the solutions consist of values of the $5$
parameters and the objective function is the corresponding value of the
$\chi^2$ from Eq.~\eqref{totalchisq}.
We define the acceptance probability function of two uncertainties, e.g.
$\chi_0$ and $\chi_1$, for two different parameter sets as
\begin{equation}
\exp\left( \frac{\chi_0 - \chi_1}{\chi_0 T} \right)
\end{equation}
where $T$ is the temperature defined in Eq.~\eqref{temperature}.
Overall, this optimization procedure proved to be the most successful.
Gradient descent was successful when the initial guess was guided
according to physical intuition. However,
simulated annealing was able to determine the global solution without this
external help.
We also investigated the robustness of this algorithm with respect to
the factor $\alpha$ used to lower the temperature.
The solution was found relatively reliably for all values of $\alpha$
between $0.85$ and $1$.
Below $\alpha = 0.85$ the process was still largely successful, but not to the
same degree.  After finding the general range of the global solution, our
simulated annealing algorithm then completes a grid search in a small region
surrounding our potential solution
to obtain the final minimum with high precision.

\section{Data Employed in the Optimization}

\label{sec:data}

As noted in \Cref{sec:computational}, all of the optimization procedures rely
on an objective function which computes how closely the \code\ output
reproduces available experimental data.
We now discuss in some detail the source and quality of the data for $^{252}$Cf(sf).
Though the available data for $^{252}$Cf(sf) are quite extensive, we still were
cautious in our selection to avoid fitting to out-of-date or low quality data.
We fit the $^{252}$Cf(sf) parameters to all eight observables mentioned in the
previous section.

We note that \cf\ is the only isotope
we consider that has data available for all observables used in the optimization.
Thus the parameters for \cf\ are the most constrained out of all the fits
performed.

The observed neutron multiplicity distribution $P(\nu)$
is taken from Ref.~\cite{SM}, an evaluation of the prompt neutron
multiplicity distributions for the spontaneous fission of a number of isotopes.
This consensus resource combines
all of the reliable direct sources of data available at this time.
We also employ this evaluation for
the full distribution $P\left( \nu \right)$, as well as the average neutron multiplicity,
$\overline\nu$, and the second and third moments, $\nu_2$ and $\nu_3$,
which are defined in Eq.~\eqref{eqn:moments}.
We have chosen to use $\overline \nu$ in addition to the actual distribution,
because the parameters in \code\ are capable of shifting $\overline \nu$ explicitly,
as well as changing the shape of $P\left( \nu \right)$, as described
in \Cref{sec:parameters}.
Note that for the optimization, we use the square root of the uncertainty given in the evaluation
since when we used the reported uncertainty, it was so low it dominated the optimization.

We have used the data from Ref.~\cite{Dushin} for the neutron multiplicity
as a function of fragment mass, $\nu(A)$.
% These data were taken at a new facility for neutron measurements.
See Ref.~\cite{Dushin} for more experimental specifics.
While Ref.~\cite{gook} also measures $\nu(A)$, these data are not used in the
optimization procedure.  We do however compare \code\ to this result in
\Cref{sec:comp_to_data}.
These two data sets are very similar, so the decision between them was largely
inconsequential in terms of the fit.

We take the neutron multiplicity as a function of TKE from Ref.~\cite{Bud}.
Since these data are from 1988, this set is rather dated.
However, this work includes a thorough and honest statistical analysis of the results
which yields reliable uncertainties.
While Ref.~\cite{gook} includes a more recent measurement of $\nu(\TKE)$,
it agrees within uncertainties with Ref.~\cite{Bud} except in regions where
the TKE is either very low or very high and is thus of low statistical significance.
We compare the \code\ results to both data sets in \Cref{sec:comp_to_data}.

The prompt fission neutron energy spectrum is taken from the
Mannhart evaluation \cite{Mannhart}.
While also somewhat dated, it is a well-established
evaluation.  We have used the non-smoothed data, which are
presented as the ratio to a Maxwellian distribution
of temperature $T  = 1.32$ \mev. We have multiplied the evaluation by the
Maxwellian at the center of each energy bin
to obtain the prompt fission neutron spectrum directly.
There is also a smoothed version of this spectrum which we have not used here
because the non-smoothed version provides an uncertainty while the smoothed
spectrum does not.
The disadvantage of using this version is the fact that there is a slight kink
around 0.1~MeV in the Mannhart spectrum.

Finally, the photon multiplicity distribution,
average photon multiplicity, and average photon energy are taken from
Ref.~\cite{Ch}, measured in 2012 using the DANCE array.
See Ref.~\cite{Ch} for more details on this analysis.

We now briefly discuss the data used in fitting the other spontaneously
fissioning isotopes in \code.  Far fewer data are available for these.
The optimizations for $^{240}$Pu(sf) and \pu{242}
were completed using the neutron multiplicity distribution,
average neutron multiplicity, average photon multiplicity,
and average photon energy.
The neutron multiplicity distribution and its moments were taken from
Ref.~\cite{SM}.  Indeed, evaluations from Ref.~\cite{SM}, available for all
spontaneously fissioning isotopes included in \code\ so far, were at times the
only data available.
The average photon multiplicity and energy for
\pu{240} and \pu{242}
both come from Ref.~\cite{Ober}.  These data, taken in 2016, are the most
recent of all the data used in the optimization.
There is also a prompt fission neutron spectrum available for \pu{240}
\cite{Alek}.  However, we have chosen not to use these data for the optimization
due not only to the limited neutron energy range but also to the
questionable quality of the data: the $^{252}$Cf(sf) neutron spectrum in
Ref.~\cite{Alek} is in disagreement with the Mannhart spectrum \cite{Mannhart}.

The neutron multiplicity distribution, average neutron multiplicity, and second
and third moments of the distribution for \cm\ are also available
from \cite{SM}.
We also fit to the neutron multiplicity as a function of fragment mass, take
from Ref.~\cite{schmidt}.
These data only have uncertainties for some values of $A$.
These uncertainties are around $0.15$,
so we took this to be the default uncertainty for the values of $\nu(A)$ without one.
We have done this because some value of uncertainty is required for the calculation of $\chi^2$.
Finally we use the neutron spectrum from Ref.~\cite{belov}.
Reference~\cite{belov} also has a neutron spectrum for \pu{242} but we
choose not to use it in the optimization due to quality issues.

The only available data for \ur\ and \pu{238} are the neutron multiplicity
distribution, the average neutron multiplicity, and the second and third
moments of the distribution from Ref.~\cite{SM}.

\section{Fit Results}
\label{sec:results}

We have confirmed the previous \cf fit results \cite{Andrew} within a reasonable margin,
produced uncertainties,
and calculated correlation matrices for the parameters.
In \Cref{table:parameters} we list our
optimized parameter values for $^{252}$Cf(sf).
These results are consistent with the default values based on physical intuition given
in \Cref{sec:parameters}.

\begin{table}
{\footnotesize
  \begin{tabular}{cCCCCC}
\hline
& e_0  & x & c & c_S & d\TKE  \\
& (\imev) &  &  &  & ({\rm MeV}) \\
\hline
\hline
$y$
&10.429
& 1.274
& 1.191
& 0.875
& 0.525
\\ \hline
$\sigma_y$ & \pm  1.090  & \pm 0.187  & \pm 0.362  & \pm 0.020  & \pm  0.078
\\ \hline \hline
$y$\cite{Andrew} & 10.37 & 1.27 & 1.18 & 0.87 & 0.52 \\ \hline
%$^{240}$Pu(sf) &
%10.750  &
% 1.307  &
% 3.176  &
% 0.908  &
%-3.219
%\\ \hline
%$\sigma$
%& \pm 0.811
%& \pm 0.155
%& \pm 0.287
%& \pm 0.206
%& \pm 2.06
%\\ \hline
\end{tabular}
}
  \caption{The optimized parameter set for $^{252}$Cf(sf)
along with the previous values of the parameters from \cite{Andrew}.}
\label{table:parameters}
\end{table}

The optimized values for $^{252}$Cf(sf)
from the preliminary optimization in Ref.~\cite{Andrew}
are shown in \cref{table:parameters}.
There is some difference between our results and those of Ref.~\cite{Andrew}
because we employ some different data sets, as well as
a slightly different optimization scheme, as described in
\Cref{sec:computational}. While some preliminary work
was also done for $^{240}$Pu(sf), we provide the first complete analysis for
this isotope, as well as the other spontaneously fissioning isotopes in \code.

We calculate the probability as a function of the $\chi^2$, as well
as the expectation values according to
\begin{align}
P( \vec{y} ) &\equiv
\left( \chi^2 ( \vec{y} ) \right)^{n / 2 - 1}
e^{-\chi^2(\vec{y})} \ ,
\label{eqn:prob}
\\
\lr{y_i} &= \int y_i\, P( \vec{y} ) \, d^5 \vec{y} \ ,
\label{eqn:expectation}
\\
\lr{y_i y_j} &= \int y_i y_j\, P ( \vec{y} ) \, d^5 \vec{y} \ ,
\label{eqn:double_expectation}
\end{align}
where $\vec{y}$ denotes the $5$-dimensional vector containing the $5$
parameter values.
We integrate over the parameter ranges.
Here $n$ is the number of degrees of freedom for all observables.
The variance and covariance of the parameters are defined as
\begin{align}
\sigma_{y_i}^2 = \lr{y_i^2} - \lr{y_i}^2&&,&&
\sigma_{y_i y_j} = \lr{y_i y_j} - \lr{y_i}\lr{y_j} \, \, .
\label{eqn:var_covar}
\end{align}
The correlation matrices in
\Cref{table:correlation_cf}
are readily calculated as
\begin{equation}
\rho_{ij} = \frac{\sigma_{y_i y_j}}{\sigma_{y_i} \sigma_{y_j}} \ .
\label{eqn:correlation}
\end{equation}
While Eqs.~\eqref{eqn:prob}-\eqref{eqn:correlation}
provide analytic definitions of these quantities,
we have numerically calculated the results in
\Cref{table:correlation_cf}
using a Hessian.
In particular, we construct a function representing the logarithm of the
probability of $\vec{y}$ and then calculate the Hessian matrix at the optimal
point using the parameter uncertainties from \Cref{table:parameters}.
The negative of the inverse of this matrix is then the covariance matrix.
We use Eq.~\eqref{eqn:correlation} to extract the correlation coefficients
displayed in the tables.
\makeatletter
\newcommand*{\bigs}[1]{{\hbox{$\left#1\vbox to35\p@{}\right.\n@space$}}}
\makeatother
\begin{table}
{\footnotesize
  \begin{comment}
\begin{equation*}
\begin{matrix}
e_0
&& x
&& c
&& c_S
&& d\TKE
\end{matrix}
\\
\begin{matrix}
\begin{matrix}
e_0 \\ x \\ c \\ c_S \\ d\TKE
\end{matrix}
&\begin{bmatrix}
1.0&-0.032&-0.737&0.924&-0.695\\
-0.032&1.0&-0.261&0.557&0.213\\
-0.737&-0.261&1.0&-0.423&0.458\\
0.924&0.557&-0.423&1.0&-0.673\\
-0.695&0.213&0.458&-0.673&1.0
\end{bmatrix}
\end{matrix}
\end{equation*}
\end{comment}
\begin{comment}
\begin{equation*}
\begin{matrix}
& e_0
& x
& c
& c_S
& d\TKE  \\
e_0 &
1.0&-0.032&-0.737&0.924&-0.695\\
x&
-0.032&1.0&-0.261&0.557&0.213\\
c &
-0.737&-0.261&1.0&-0.423&0.458\\
c_S &
0.924&0.557&-0.423&1.0&-0.673\\
d\TKE &
-0.695&0.213&0.458&-0.673&1.0
\end{matrix}
\end{equation*}
\end{comment}
\begin{tabular}{CCCCCC}
& e_0 & x & c & c_S & d\TKE \\ \ \\
\begin{matrix}
e_0 \\ x \\ c \\ c_S \\ d\TKE
\end{matrix}
&
\bigs{\{ }
\begin{matrix}
1.0 \\ -0.032 \\ -0.737 \\ 0.924 \\ -0.695
\end{matrix}
&
\begin{matrix}
-0.032 \\ 1.0 \\ -0.261 \\ 0.557 \\ 0.213
\end{matrix}
&
\begin{matrix}
-0.737 \\ -0.261 \\ 1.0 \\ -0.423 \\ 0.458
\end{matrix}
&
\begin{matrix}
0.924 \\ 0.557 \\ -0.423 \\ 1.0 \\ -0.673
\end{matrix}
&
\begin{matrix}
-0.695 \\ 0.213 \\ 0.458 \\ -0.673 \\ 1.0
\end{matrix}
\bigs{\}}
\\
\end{tabular}
}
\caption{
Correlation coefficients for \cf.
}
\label{table:correlation_cf}
\end{table}
\begin{comment}
%\begin{table}
%\begin{tabular}{cCCCCC}
%\hline
%$^{240}$\textbf{Pu(sf)}
%&e_0 & x & c & c_S & d\TKE \\
%\hline \hline
%$e_0$ & 1 &&&&
%\\ \hline
%$x$ & -0.332 & 1 &&&
%\\ \hline
%$c$ & 0.174 & -0.058 & 1 &&
%\\ \hline
%$c_S$ & -0.918 & 2.292 \cdot 10^{-4} & 0.1924 & 1 &
%\\ \hline
%$d$TKE & -0.973 & 0.332 & -0.173 & 3.0044 & 1
%\\ \hline
%\end{tabular}
%\caption{
%Correlation coefficients for \pu{240}.
%}
%\label{table:correlation_pu}
%\end{table}
\end{comment}

In \Cref{fig:contours_cf}
we present contour plots of the $\chi^2$ for different values of
the parameters. We vary the linear combinations listed on the axes and fix all
parameters which are not listed to their central values.
These plots can be interpreted as surfaces in the higher dimensional
space which gives us a particular uncertainty for any choice
of $5$ parameters. As previously described,
the particular linear combinations of parameters was based on physical intuition
in order to best illustrate the nature of the parameter space we are working in.
The linear combination of parameters on the axes in \Cref{fig:contours_cf}
was determined by eye according to the contour plots of the individual parameters.
We do not show the variance as an error bar in the lower plot, because it effectively fills
the entire displayed range.

As discussed in greater detail in \Cref{sec:computational},
employing a grid search will always find the proper solution by testing
every possible combination, whereas the alternative optimization methods
attempt to ``climb'' around these contours in order to find
the point of minimum uncertainty.
These plots show that there is not always a clear ``valley'' of
minimal uncertainty, and a simple grid approach is very likely to fall into a local minimum.

\begin{comment}
\begin{figure}[t]
\centering
\includegraphics[width=0.49\textwidth]{Pu240/contours/well.pdf}
\\
\includegraphics[width=0.49\textwidth]{Pu240/contours/well_closeup.pdf}
\caption{(Color online) (a) Contour plot of $\chi^2$ relative to
various parameter values for $^{240}$Pu(sf) for the full range of tested parameters.
(b) Contour zoomed in to focus around optimized parameter values.}
\label{fig:contours_pu}
\end{figure}
\end{comment}

It is worth addressing the size of the $\chi^2$ in our results.
Our $\chi^2$ is summed over the uncertainty estimate
for each bin of the data sets and evaluations we fit to.
While this value is very large,
this should not suggest that our fit is low quality, see \Cref{sec:computational}.
\begin{comment}
The contours in \Cref{fig:contours_cf}
show particularly high $\chi^2$ values, because these
plots show the uncertainty over the full range of the parameters.
As such, the edges of these plots correspond to values of the parameters which
are far from a physically-reasonable result.
These high uncertainties confirm that the outer regions are indeed
undesirable.
\end{comment}

While our main focus is on \cf,
we also completed the same analysis for
\ur, \pu{238}, \pu{240}, \pu{242}, and \cm, the other spontaneously
fissioning isotopes currently included in \code\ \cite{CPC_NVA}.
These results are listed in \Cref{table:all_parameters}.
A comparison of the fits for these isotopes to the data and evaluations used in the fits
are available in the Appendix.

While we have determined uncertainties on the parameter values for these
isotopes, obtaining reliable correlation matrices for them
is difficult. \Cref{table:all_parameters}, which also lists the number of data
sets and evaluations used in our fits, makes this obvious.
If only a single evaluation is available, as is the case for \ur\ and \pu{238},
it is difficult to say, without other constraints, how changing one parameter
with respect to the others would affect the correlation.

\begin{table*}
\centering
\begin{tabular}{|c|C|C|C|C|C|c|c|}
\hline
 & e_0 \; (/{\rm MeV}) & x & c & c_S & d\TKE \; ({\rm MeV})
& \text{\# Data Sets}
& \text{\# Evaluations}\\
\hline
\multicolumn{8}{c}{\ur } \\ \hline
$y$ & 10.391 & 1.220 & 0.939 & 0.899 & -1.375 & 0 & 1 \cite{SM}
\\ \hline
$\sigma_y$ & \pm 0.352  & \pm 0.071  & \pm 0.283  & \pm 0.280  & \pm 0.727 &-&-
\\ \hline
\multicolumn{8}{c}{\pu{238} } \\ \hline
$y$ & 10.521  & 1.232  & 1.968  & 0.893 & -1.408 & 0 & 1 \cite{SM}
\\ \hline
$\sigma_y$ & \pm  0.581  & \pm 0.221  & \pm 0.071  & \pm 0.071  & \pm 3.424  &-&-
\\ \hline
\multicolumn{8}{c}{\pu{240} } \\ \hline
$y$ & 10.750  & 1.307  & 3.176  & 0.908 & -3.219 & 1 \cite{Ober} & 1 \cite{SM}
\\ \hline
$\sigma_y$ & \pm 0.138  & \pm 0.071  & \pm 0.355  & \pm 0.023 & \pm 0.112  &-&-
\\ \hline
\multicolumn{8}{c}{\pu{242} } \\ \hline
$y$ & 10.018  & 1.144  & 3.422  & 0.911 & -1.662 & 1 \cite{Ober} & 1 \cite{SM}
\\ \hline
$\sigma_y$ & \pm  1.768  & \pm 0.152  & \pm 0.341  & \pm 0.257  & \pm 0.118  &- &-
\\ \hline
\multicolumn{8}{c}{\cm } \\ \hline
$y$ & 10.488  & 1.239  & 1.391  & 0.906 & -4.494 & 2 \cite{schmidt,belov}
& 1 \cite{SM} \\ \hline
$\sigma_y$ & \pm 1.519  & \pm 0.148  & \pm 0.582  & \pm 0.322  & \pm 0.167  &-&-
\\ \hline
\multicolumn{8}{c}{\cf } \\ \hline
$y$ &10.429 & 1.274 & 1.191 & 0.875 & 0.525 & 4 \cite{Dushin,gook,Bud,Ch}
& 2 \cite{SM,Mannhart}
\\ \hline
$\sigma_y$ & \pm  1.090  & \pm 0.187  & \pm 0.362  & \pm 0.020  & \pm  0.078  &-&-
\\ \hline
\end{tabular}
\caption{Results of the optimization for all spontaneously-fissioning isotopes
modeled by \code.  The best fit values of the five parameters, $y$,
and their associated standard deviations, $\sigma_y$, are given for each isotope.  In
addition, the number of data sets and evaluations used for each isotope are
indicated, along with the references for these data.  Note that in the case
of Ref.~\cite{SM}, the evaluation gives the result for multiple observables:
$P(\nu)$, $\overline \nu$, $\nu_2$ and $\nu_3$.}
\label{table:all_parameters}
\end{table*}

\section{Interpretation}

\label{sec:interpretation}

In this section we develop a physical interpretation for the parameter values obtained in
\Cref{sec:results}.
We have treated all spontaneously-fissioning isotopes in \code\
individually, with all five parameters allowed to
vary independently regardless of how many data sets are available to constrain
them.  This is not unreasonable because we do not generally expect the parameters to have
the same value for all fissioning systems.

The parameters $c$
and $d$TKE, which influence $P(\nu)$ and $\overline \nu$, are perhaps best
constrained because evaluations of $P(\nu)$ and the values of its moments are available
for all isotopes in \code.  Indeed, for some cases, these are the only available
parameter constraints.  Because the shape of  $P(\nu)$ and its moments affect
both $c$ and $d$TKE, and given that $\overline \nu$, $\nu_2$ and $\nu_3$ vary
considerably from isotope to isotope, we can expect $c$ and $d$TKE to vary
independently as well.
We might expect the largest range of variation for these as well.

The other three parameters ($e_0$, $x$ and $c_S$) have fewer data available to
constrain them.
The average photon multiplicity and energy per photon can be used to guide the value of
$c_S$ for \pu{240}, \pu{242}, and \cf.
We have $\nu(A)$ data to constrain
$x$ for \cm\ and \cf.  Finally, we have used spectral data
for \cm\ and \cf\ which provides a partial constraint on $e_0$.  We remark that
it is only partial because all parameters influence the prompt fission neutron
spectrum.
While $e_0$ is directly related to the temperature, see Eq. \eqref{Eq:scission}
and thus the slope of the prompt fission neutron spectrum,
the parameters $c_S$, $x$, and $c$ are also related to the temperature, at least
indirectly. Recall that $c$ sets the level of thermal fluctuations,
Eq. \eqref{eqn:c};
$x$ controls the sharing of excitation energy between fragments,
initially related to the level density, Eq. \eqref{eqn:x}; and $c_S$
is related to the scission temperature, Eq. \eqref{eqn:c_S}.
Thus these parameters all also influence the spectrum.

We expect $c_S$ to be less than unity while we expect $x$ and $c$ to
be larger than unity.  Since $e_0$ is related to the level density parameter,
we expect a value of $8-12$~\imev\ from other work \cite{RIPL3}.
We may also expect $d$TKE to vary
considerably to make up for a lack of other constraints on the parameters aside
from $\overline \nu$ and also, because, on some cases the input data used for
TKE$(A_H)$ have either large uncertainties based on low-statistical samples,
or no uncertainties given.  An examination of the results in
\Cref{table:all_parameters} can give us insight into how well the
optimization procedure met our expectations.

The parameter values for the spontaneously fissioning
isotopes in \code\ 2.0.2 \cite{CPC_NVA} were obtained in a far more empirical
fashion.  The values for \cf\ were taken from Ref.~\cite{Andrew}, obtained by
a grid search procedure.  Universal values were then assumed for $c_S$ and
$e_0$.  (We note that while $e_0$ was fixed to the \cf\ value from
Ref.~\cite{Andrew} for neutron induced fission, the value
$e_0 = 10.0724$/MeV was retained from \code\ version 1.0 \cite{CPC}
for the other spontaneously fissioning isotopes.)  While one can reasonably
assume that $e_0$ has a universal value since the nuclear level
densities are related to nuclear structure and not reaction dependent,
$c_S$ was fixed for expedience.  The
$x$ parameter for \pu{240} in \code\ 2.0.2 was fixed from experimental analysis
of neutron-neutron correlations in Ref.~\cite{jerome_angular}, an observable not
used in this optimization because it requires full analysis of the detector
setup in each case.
However, these correlations exhibit strong sensitivity to $x$
\cite{RVJR_corr}.  For other spontaneously fissioning isotopes, it was taken
to be $\sim 1.2$.

The parameter $c$ was fixed via examination of $P(\nu)$.  Finally,
$d$TKE was tuned to $\overline \nu$ after the other parameter values were fixed.
The work in this paper is the first to make a full optimization of all
parameters for all isotopes.  It is interesting to compare how well this
empirical approach compares with the numerical optimization performed in the
current paper.

As already noted, we do not expect $c$ and $d\TKE$ to be independent of
isotope.
As can be seen in \Cref{table:all_parameters}, they are not.
The values obtained for $c$ are driven entirely by $P(\nu)$
and its moments.
In the cases where $c$ is large, $c>3$,
\pu{240} and \pu{242}, it is because despite the low average neutron
multiplicity, $P(\nu)$ is broader than might be expected for low
$\overline \nu$.  In such cases, the range of $c$ needs to be increased to
match the higher moments of the multiplicity distribution, $\nu_2$ and $\nu_3$.
There is also one exception to the expectation that $c\geq 1$, \ur.
In this case, the evaluated $P\left( \nu \right)$ is actually more peaked than a
distribution with $c = 1$ for the same $\overline \nu$, requiring the
fluctuations to be reduced to achieve agreement with the evaluated
$P\left( \nu \right)$ and its moments.
Note, however, that, within uncertainties, $c$ is still compatible with unity
in this case. The values of $c$ in \code\ 2.0.2 were $0.92$, $1.91$, $3$, $3.4$ and
$1.34$ for \ur, \pu{238}, \pu{240}, \pu{242} and \cm\ respectively, in addition
to the value of 1.18 found in Ref.~\cite{Andrew} for \cf.  These empirical
guesses are very close to the results obtained from our current optimization
based on evaluations that explicitly constrain $c$.

Next, as indicated, we expect $d\TKE$ to vary from case to case,
independent of isotope.
Ideally $d\TKE$ should be zero with a perfect model along with high statistics
input yields and $\TKE\left( A_H \right)$.
This is indeed the case for \cf, a well-measured standard with a high
spontaneous fission rate. $d\TKE$ is small for \cf: $d\TKE = 0.525$~MeV
here and 0.52~MeV in Ref.~\cite{Andrew}.
We now compare our optimized $d\TKE$ values for the other isotopes studied here
with those in \code\ 2.0.2 \cite{CPC_NVA}: $d\TKE = -1.345$~MeV, $-1.366$~MeV,
$-3.071$~MeV, $-1.600$~MeV, and $-4.35$~MeV for \ur, \pu{238}, \pu{240},
\pu{242} and \cm\ respectively.
These values are in rather good agreement with those found in our optimization.
We note that the large range in $d\TKE$ values is expected
and does not affect the physical interpretation of the parameters.

It is notable that these values are, in contrast to that for \cf, all negative
and the absolute values are considerably larger.
A negative value for $d\TKE$ indicates that the
reported TKE$(A_H)$ distribution is too high, reducing the overall available
excitation energy for neutron emission.
However, the measured fission rates are much lower for other
isotopes and large fluctuations exist in the data.
In most of these cases, the number
of fission events measured was small so that not many events go into each $A_H$ bin.
\code\ samples the yields and $\TKE\left( A_H \right)$ directly
from measured fission fragment data, often with undefined or
unquantified systematic uncertainties.
Thus the input TKE$(A_H)$ in \code\ in these cases are based on low
statistics, sometimes without uncertainties on the data, and with unknown
systematic errors.  Introducing $d\TKE$ is a way to correct
for these unknowns as well as offering a means to compensate for any remaining,
unquantified, physics effects.

Previously, the value of $c_S$ was fixed at $0.87$
for every isotope of \code.
We can see that our results for \cf\ agree with this but the \pu{240}\
result is somewhat larger.
We generally find that the spin temperature is close to the scission temperature,
resulting in fragment spins close to the maximum available rotation.
We note that there is some correlation between $c_S$ and $d\TKE$.  While it
may be especially weak for \cf, it could be responsible for the differences
observed between the values of $d\TKE$ in \code\ 2.0.2 \cite{2.0.2UserManual} and
\Cref{table:all_parameters} since changing $c_S$ changes $E_{\rm rot}$ which,
in turn, modifies $E_{\rm stat}$, thus ultimately affecting $d\TKE$.  Increasing
$c_S$, as for {\it e.g.} \pu{240} to 0.908 from 0.87, increases $E_{\rm rot}$.
Thus for a fixed scission energy $E_{\rm sc}$ then, $E_{\rm stat}$ is reduced,
decreasing the energy available for neutron emission.  To keep $\overline \nu$
fixed, absent other variation, $d\TKE$ has to decrease.  This is seen in
\Cref{table:all_parameters} as $d\TKE$ is now $-3.219$~MeV instead of
$-3.07$~MeV.  Similar reductions of $d\TKE$ can be seen for increased $c_S$
in the other cases studied.

The parameter $x$ controls the distribution of statistical excitation
energy between the fragments after scission.
It is well established that $x$ is greater than unity based on $\nu(A)$ data
and previous measurements of the average neutron multiplicities from the light
and heavy fragments respectively.  The previous values of $x$
\cite{2.0.2UserManual} generally assumed $x \sim 1.2$ aside from the value of
1.27 established in Ref.~\cite{Andrew} for \cf\ and the $1.3$ found for \pu{240} based
neutron-neutron angular correlation data \cite{jerome_angular}.  The estimates
of $x \sim 1.1-1.2$ for the other spontaneously fissioning isotopes were borne
out by our independent fits.  Despite the fact that the $x$ range was
$1 < x <1.5$ in all the fits, with $\nu(A)$ data only available for \cm\ and \cf,
the optimized values are very similar to the default of $x \sim 1.2$ assumed
previously.

The values of $e_0$ in \Cref{table:all_parameters} are remarkably similar,
between $10$\imev\ and $10.75$\imev\ for all isotopes, despite the wide range,
$7 < e_0 < 12$\imev.  This is particularly striking
because, of the observables considered, only the prompt fission neutron
spectrum shows any direct dependence on $e_0$, even though it also depends on every other
parameter.  For example, increasing $c_S$ gives more available excitation energy
to neutron emission which could, in principle, increase the average energy per
neutron rather than increasing the number of neutrons and thus change the slope
of the prompt fission neutron spectrum.  Giving a larger share of the excitation
energy to the light fragment would also influence the average neutron energy and
thus the spectral shape, as would modifying the fluctuations in excitation
energy via a change in $c$ because increased fluctuations in statistical
excitation, while modifying $P(\nu)$, can also modify the neutron energy.
Despite this, $e_0$ remains remarkably similar for all cases, even though it
was fit independently.  Recall that, even though
Eq.~(\ref{Eq:scission})
refers to the temperature at scission, the same equation applies to
every neutron emitted throughout the fission process as well.  Thus, in turn,
$e_0$ influences the emission of every neutron even though it only has a visible
influence on the neutron spectrum and not on {\it e.g.} $\nu(A)$.

It is worth noting, that the results for $e_0$, $x$, and $x_S$
in \Cref{table:all_parameters} are all very similar.
The similarity of these parameters is an indication that the
mechanisms employed in \code\ are physically relevant.
As mentioned in \Cref{sec:parameters}, we allow a larger variation of $c$ for
some isotopes because the widths of the neutron multiplicity distributions can
vary significantly between isotopes.

We can also gain physical intuition from the results in
\Cref{table:correlation_cf}.
As expected mathematically, these matrices are symmetric with unity along the
diagonal.  In addition, the off-diagonal values are bounded by unity.
A positive correlation between two parameters suggests that, when
one of these parameters is raised, in order to maintain agreement with the
data, the other must increase as well.  A negative correlation suggests that
when one is raised, the other needs to decrease to compensate.
A correlation with an absolute value close to unity indicates that the
relationship between parameters is strong while, when the correlation is close
to zero, this relationship is weak.

\begin{comment}
We see that there is a weak correlation between $x$ and $c_S$ and a similarly
weak correlation between $c$ and $c_S$.
Because, according to the physical interpretation of these parameters,
$c_S$ controls fragment spin, and therefore photon emissions,
while $x$ and $c$ are related to the thermal excitation of the fragments and
therefore neutron emissions, the relatively weak correlations reflect the
independence of these parameters with respect to each other.
We also note that there is a relatively weak correlation between $x$ and $c$.
Since $x$ adjusts how the statistical excitation energy is divided between the
fragments, it primarily affects $\nu\left( A \right)$, whereas $c$
adjusts the width of the multiplicity distribution $P\left( \nu \right)$.
This is reflected in the weak correlation between these parameters.
\end{comment}

We now discuss the correlations between the input parameters, starting with the correlation of $e_0$ with the other parameters and proceeding across \Cref{table:correlation_cf}.
The correlation between $e_0$ and $x$ is the weakest.  This is because the initial excitation energy partition is divided up between the two fragments according to their level densities, as described in \Cref{sec:parameters}.  The ratio of the level densities is independent of changes to $e_0$.  The parameter $x$ is a perturbation on that ratio, resulting in a weak correlation.  The correlation between $e_0$ and $c$ is large and negative so that, when $e_0$ is increased, $c$ decreases.  Since $e_0$ is related to the fragment energy before neutron emission, see  Eq.~(2), increasing $e_0$ while keeping the energy for neutron emission fixed forces the temperature to increase.  Since $c$ is related to the thermal fluctuations in the decaying nucleus, if the temperature increases, then the fluctuations can also increase so that $c$ has to decrease to compensate.  The correlation between $e_0$ and $c_S$ is the strongest of all, near $+1$, implying that $c_S$ must increase when $e_0$ increases.  Again, increasing $e_0$ can imply an increase in temperature and a probability of greater neutron emission.  To compensate, $c_S$ needs to increase to give more rotational energy to the fragments and more photon emission to keep the neutron emission fixed.  There is also a relatively strong, negative, correlation between $e_0$ and $d$TKE.  A higher fragment temperature could either lead to increased neutron emission or emission of higher energy neutrons.  If fewer neutrons are emitted with higher average energy, then dTKE would need to decrease to compensate to increase the total excitation energy to increase neutron emission.

There is a relatively weak correlation between $x$ and $c$.  Since $x$ adjusts how the statistical excitation energy is divided between the fragments, it primarily affects $\nu(A)$ whereas $c$ adjusts the width of the multiplicity distribution $P(\nu)$.  A moderate positive correlation is seen between $x$ and $c_S$.  If $x$ is increased to give more energy to the light fragment, the average neutron multiplicity can be expected to increase.  Thus $c_S$ must increase to take more rotational energy and keep the neutron multiplicity constant.  The correlation between $x$ and $d$TKE is small and positive so that, if $x$ increases neutron emission, then $d$TKE must increase to compensate and reduce the total excitation energy.

A moderate, negative correlation is seen between $c$ and $c_S$.  If $c$ increases, $\overline \nu$ will decrease so that, for $\overline \nu$ to be maintained, the rotational energy, and thus $c_S$, has to decrease.  On the other hand, the correlation between $c$ and $d$TKE is moderate but positive.  If neutron emission increases with increasing $c$, then to increase the total excitation energy to compensate, $d$TKE has to increase also to decrease the neutron multiplicity.  Finally, there is a relatively large negative correlation between $c_S$ and $d$TKE.  If $c_S$ is increased, the fragment spin and thus rotational energy increases, taking energy away from that available for statistical neutron emission.  Thus $d$TKE has to decrease to give more total excitation energy to the fragments and maintain the value of $\overline \nu$.

\section{Comparison to Data}
\label{sec:comp_to_data}

\begin{figure}[t]
\centering
\includegraphics[width=0.49\textwidth]{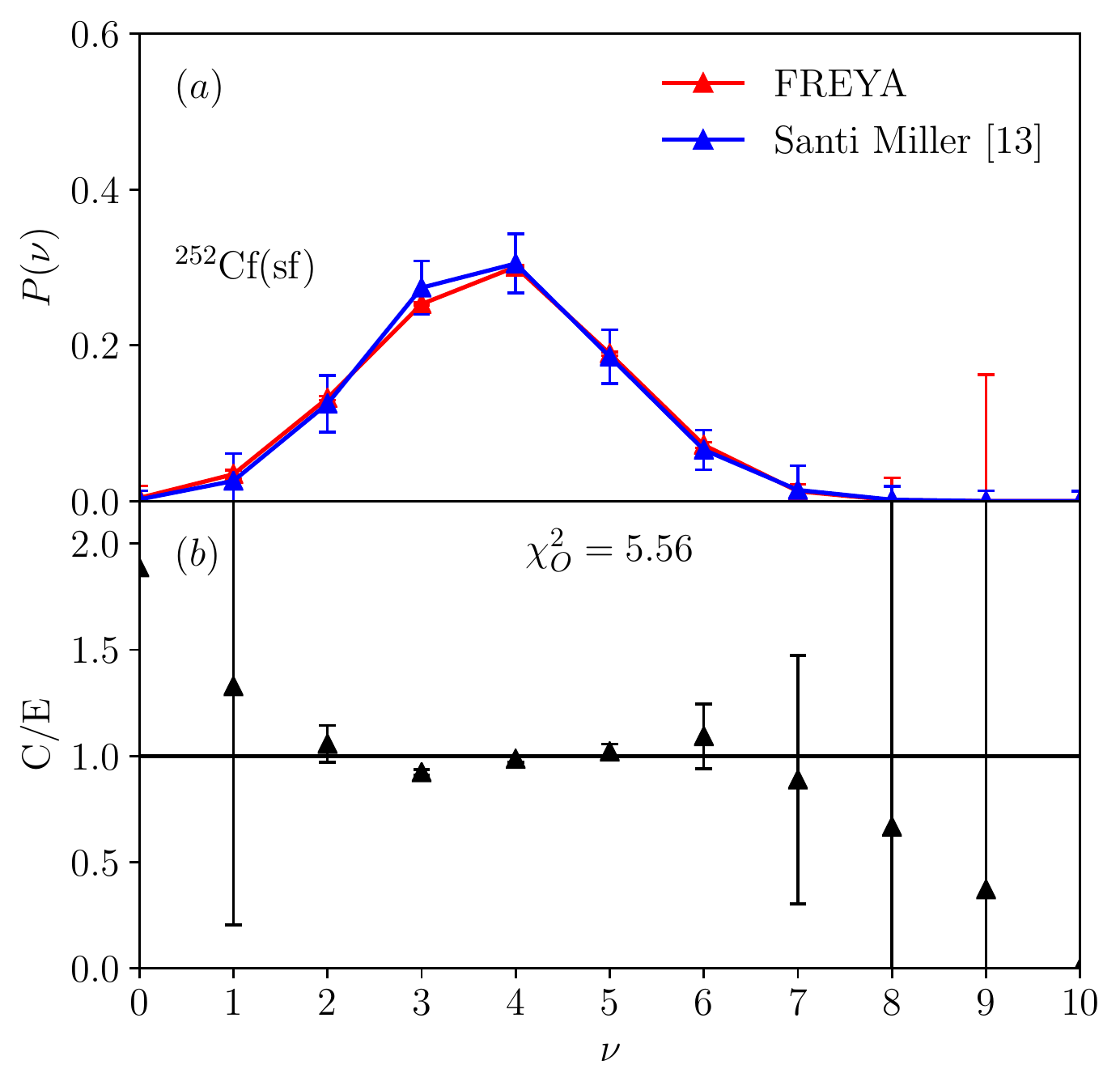}
\caption{(Color online) (a) Neutron multiplicity distribution for \cf\
compared to the Santi-Miller evaluation Ref.~\cite{SM}.
Note that for the comparison, we use the square root of the uncertainty given in the evaluation
since the reported uncertainty was so low it dominated the optimization.
Note that the uncertainty on the \code\ calculation is the variance of the result,
which we then present as a standard deviation.
Therefore this should not be interpreted as the range of values we should expect from \code.
In this particular case, since we are not calculating an average, we instead take this variance to be
$1 / \sqrt{N}$ where $N$ is the number of observed events in that bin.
(b) Ratio of calculated values to evaluation results.}
\label{fig:cf_p_nu}
\end{figure}

\begin{figure}[t]
\centering
\includegraphics[width=0.49\textwidth]{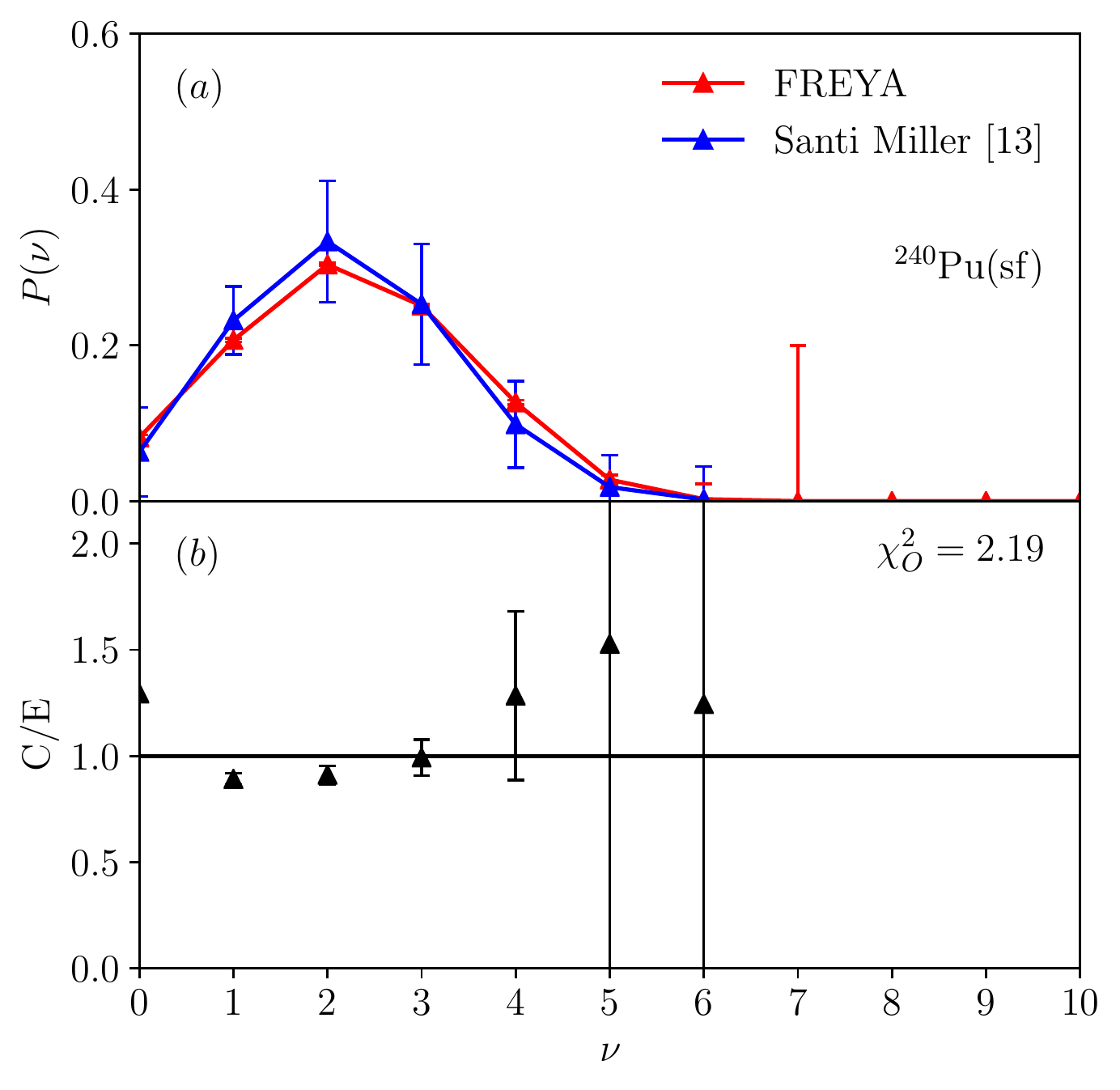}
\caption{(Color online)
The same result as in Fig.~\ref{fig:cf_p_nu} for \pu{240}.}
\label{fig:pu240_p_nu}
\end{figure}

\begin{figure}[t]
\centering
\includegraphics[width=0.49\textwidth]{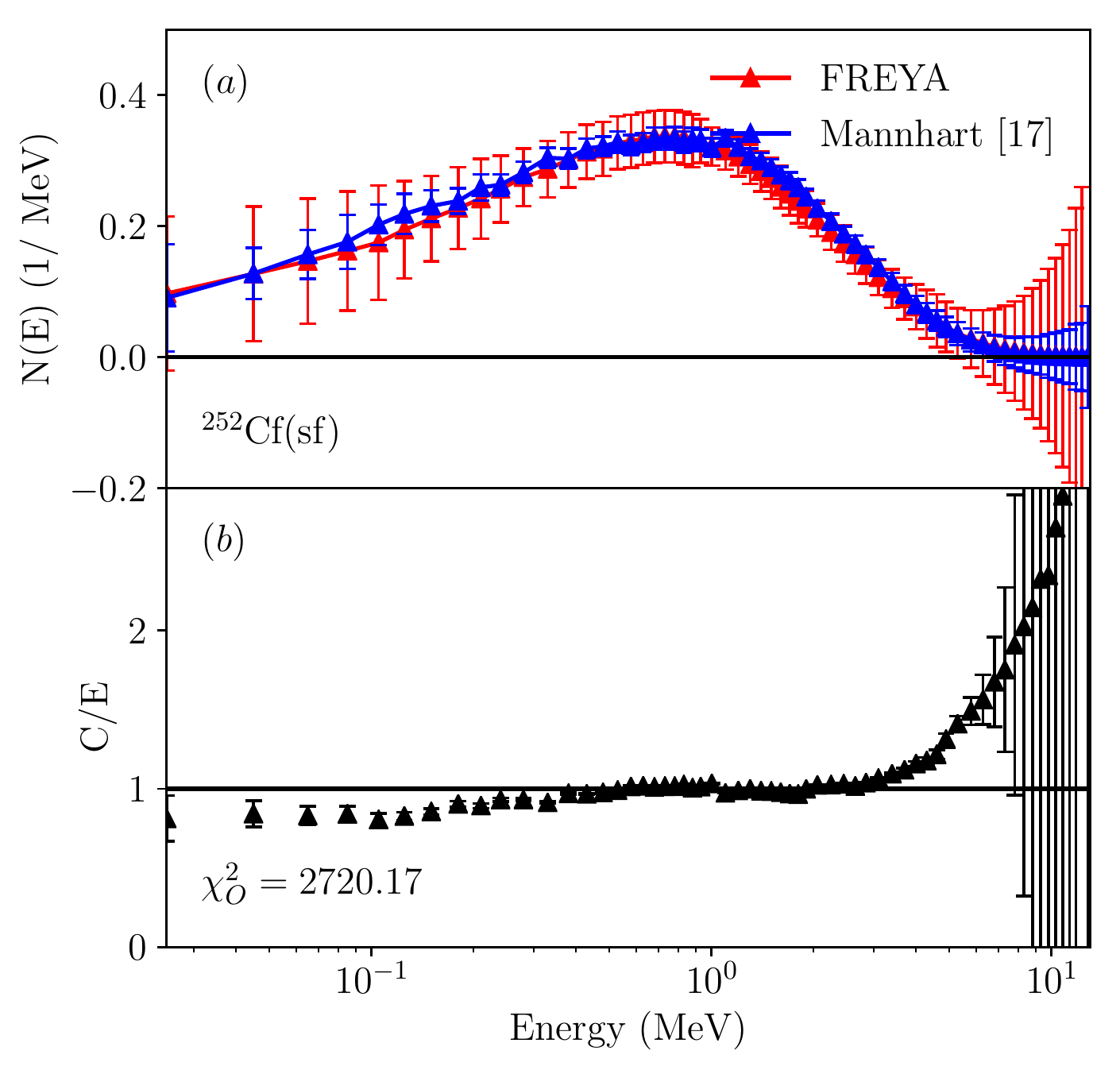}
\includegraphics[width=0.49\textwidth]{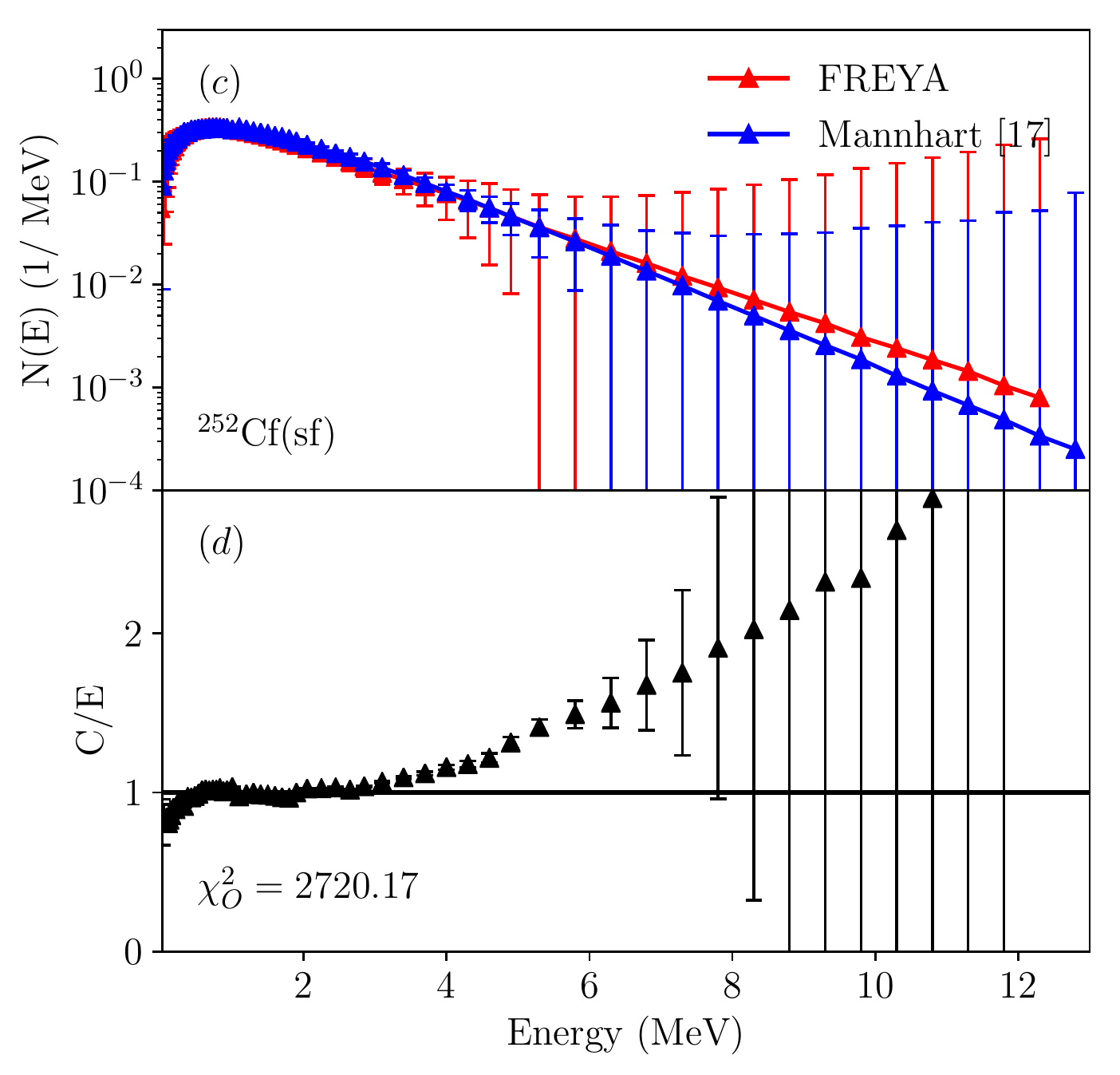}
\caption{(Color online) (a) Neutron energy spectrum for \cf\ from \code\
  as well as the Mannhart evaluation \cite{Mannhart}
with a logarithmic scale on the $x$ axis.
(b) Ratio of calculated values to evaluation results.
A logarithmic scale is used on the $x$-axis in both cases.
In (c) and (d) the same results are shown
now with a logarithmic scale on the $y$-axis, and a linear scale on the $x$-axis
in order to highlight the difference and uncertainty at high energy.}
\label{fig:mannhart}
\end{figure}

We now use the optimized parameters presented in \Cref{sec:results} to generate
a set of one million \code\ events,
and compare this to the data used in our optimization, along
with some data which were not included.
We present the direct comparisons as well as ratios of the calculated
to experimental values (C/E).
It is important to note that throughout the section, any uncertainties
given on the results from \code\ arise from calculating the variance arising from the propagation of the uncertainties on the model parameters and are not indicative of any statistical uncertainty in the \code\ calculation.
In cases where there is no relevant variance to calculate,
we instead use $1/\sqrt{N}$, where $N$ is the relevant event multiplicity in the
bin of a distribution.
Especially in this case, this 'uncertainty' should not be compared to the uncertainty
on the experimental data.

\begin{table}
  \footnotesize{
\begin{tabular}{CCCC}
\hline
\text{ $\nu_n$ } & \text{ Evaluation } &\text{\code } & \text{C/E}\\
\hline \hline
\multicolumn{4}{c}{\ur }
\\ \hline
\overline\nu &1.98\pm 0.03&2.0\pm 0.94&1.01\pm 0.22\\
\nu_2 &2.8743\pm 0.1411&2.87\pm 3.37&1.0\pm 1.37\\
\nu_3 &2.8219\pm 0.481&2.83\pm 9.81&1.0\pm 11.71\\
\hline\hline
\multicolumn{4}{c}{\pu{238} }
\\ \hline
\overline\nu &2.19\pm 0.07&2.17\pm 1.15&0.99\pm 0.27\\
\nu_2 &3.8736\footnote{No uncertainty reported.} &3.85\pm 4.35&0.99\pm 1.26\\
\nu_3 &5.417\footnote{No uncertainty reported.} &5.25\pm 10.97&0.97\pm 4.1\\
\hline\hline
\multicolumn{4}{c}{\pu{240} }
\\ \hline
\overline\nu &2.154\pm 0.005&2.22\pm 1.25&1.03\pm 0.33\\
\nu_2 &3.7889\pm 0.029&4.26\pm 4.88&1.12\pm 1.66\\
\nu_3 &5.2105\pm 0.1492&6.53\pm 13.3&1.25\pm 6.51\\
\hline\hline
\multicolumn{4}{c}{\pu{242}}
\\ \hline
\overline\nu &2.149\pm 0.008&2.12\pm 1.19&0.99\pm 0.3\\
\nu_2 &3.8087\pm 0.036&3.79\pm 4.51&0.99\pm 1.4\\
\nu_3 &5.3487\pm 0.036&5.36\pm 12.13&1.0\pm 5.14\\
\hline\hline
\multicolumn{4}{c}{\cm }
\\ \hline
\overline\nu &2.71\pm 0.01&2.7\pm 1.16&1.0\pm 0.18\\
\nu_2 &5.941\pm 0.0188&5.95\pm 5.46&1.0\pm 0.84\\
\nu_3 &10.112\pm 0.175&10.17\pm 16.78&1.01\pm 2.75\\
\hline\hline
\multicolumn{4}{c}{\cf }
\\ \hline
\overline\nu &3.757\pm 0.01&3.74\pm 1.3&1.0\pm 0.12\\
\nu_2 &11.9517\pm 0.0188&11.94\pm 8.79&1.0\pm 0.54\\
\nu_3 &31.668\pm 0.175&31.84\pm 39.94&1.01\pm 1.59\\
\hline
\end{tabular}
}
\caption{Average neutron multiplicity and the second and third moments
of the neutron multiplicity distribution for all six isotopes in \code\
compared with the evaluations in Ref.~\cite{SM}.
Note that the uncertainty on the \code\ calculation is a calculation of the variance of the result,
and should therefore not be interpreted as the range of values we should expect \code\ to return.
}
\label{table:nubar}
\end{table}

\begin{table}
\footnotesize{
  \begin{tabular}{CCCC}
\hline
\text{Average} & \text{ Measured } &\text{\code } & \text{C/E} \\
\hline \hline
\multicolumn{4}{c}{\ur }\\
\hline
\overline{N}_{\gamma} & - & 6.49\pm 2.42 & - \\
\overline{\epsilon_\gamma} \, \text{(MeV)}& - &0.94\pm 0.87& - \\
\hline\hline
\multicolumn{4}{c}{\pu{238} }\\
\hline
\overline{N}_{\gamma} & - &6.47\pm 2.43& - \\
\overline{\epsilon_\gamma} \, \text{(MeV)}& - &1.05\pm 0.93& - \\
\hline\hline
\multicolumn{4}{c}{\pu{240} }\\
\hline
\overline{N}_{\gamma}\text{\cite{Ober}} &8.2\pm 0.4&6.6\pm 2.48&0.8\pm 0.09\\
\overline{\epsilon_\gamma}\text{\cite{Ober}} \, \text{(MeV)}&0.8\pm 0.07&1.0\pm 0.91&1.24\pm 1.26\\
\hline\hline
\multicolumn{4}{c}{\pu{242} }\\
\hline
\overline{N}_{\gamma}\text{\cite{Ober}} &6.72\pm 0.07&6.61\pm 2.43&0.98\pm 0.13\\
\overline{\epsilon_\gamma}\text{\cite{Ober}} \, \text{(MeV)}
&0.843\pm 0.012&0.96\pm 0.89&1.14\pm 1.12\\
\hline \hline
\multicolumn{4}{c}{\cm }\\
\hline
\overline{N}_{\gamma} & - &7.07\pm 2.56& - \\
\overline{\epsilon_\gamma} \, \text{(MeV)}& - &1.01\pm 0.93& - \\
\hline\hline
\multicolumn{4}{c}{\cf }\\
\hline
\overline{N}_{\gamma}\text{\cite{Ch}} &8.14\pm 0.4&7.71\pm 2.8&0.95\pm 0.12\\
\overline{\epsilon_\gamma}\text{\cite{Ch}} \, \text{(MeV)}&0.94\pm 0.05&0.91\pm 0.86
&0.97\pm 0.83\\ \hline
\end{tabular}
}
  \caption{Average photon multiplicity and average energy per photon for all six
  isotopes in \code\ compared with experimental data (when available).
  The data for \cf\ comes from Ref.~\cite{Ch} while the data for \pu{240} and
  \pu{242} both come from Ref.~\cite{Ober}.
Note that the uncertainty on the \code\ calculation is a calculation of the variance of the result,
and should therefore not be interpreted as the range of values we should expect \code\ to return.
}
\label{table:gammabar}
\end{table}

As can be seen in \Cref{fig:cf_p_nu,fig:pu240_p_nu} we reproduce the
neutron probability distribution $P(\nu)$ within very low uncertainty
in both cases.
As discussed in \Cref{sec:data}, the neutron multiplicity distributions
are well established for both of these isotopes.
Even where we do differ from the result, the uncertainties on C/E are still
compatible with unity.
We also reproduce the average neutron multiplicity
in \Cref{table:nubar} to one or two decimal points, effectively the regime
in which we can accurately interpret the \code\ results.
We also note that the neutron multiplicity moments have improved
with the new set of parameters over those from Ref.~\cite{Andrew}.  The moments
are important for criticality studies.
Note that the uncertainty on the \code\ calculation is a calculation of the variance of the result,
and should therefore not be interpreted as the range of values we should expect \code\ to return.

The prompt fission neutron spectrum is compared to the Mannhart evaluation
in \Cref{fig:mannhart}. We show the results on a logarithmic scale on the $x$
axis in (a) and (b) as well as the $y$ axis in (c) and (d). These particular
scales allows us to get a good sense of the behavior of the spectrum in
both the low energy regime from $0$ to $1$ MeV, in (a) and (b), as well as the
high energy range from $1$ to $12$ MeV, in (c) and (d).
\code\ reproduces this
distribution with high accuracy in the low energy range except for the slight
kink around $0.1$~MeV present in the non-smoothed version of the Mannhart
evaluation, which we employ because it includes uncertainties.
There is a more significant deviation in the high energy range,
but the range of uncertainty in C/E is consistent with unity for
neutron energies above $7$~MeV.
It is also important to note that the uncertainties are extremely large in
this high-energy region for both the experimental data and the \code\ output.
We note that the \code\ uncertainties in the high energy tail of the spectrum
can be reduced by generating a larger number of events while the uncertainties
on the evaluation cannot.

\begin{figure}[t]
\centering
\includegraphics[width=0.49\textwidth]{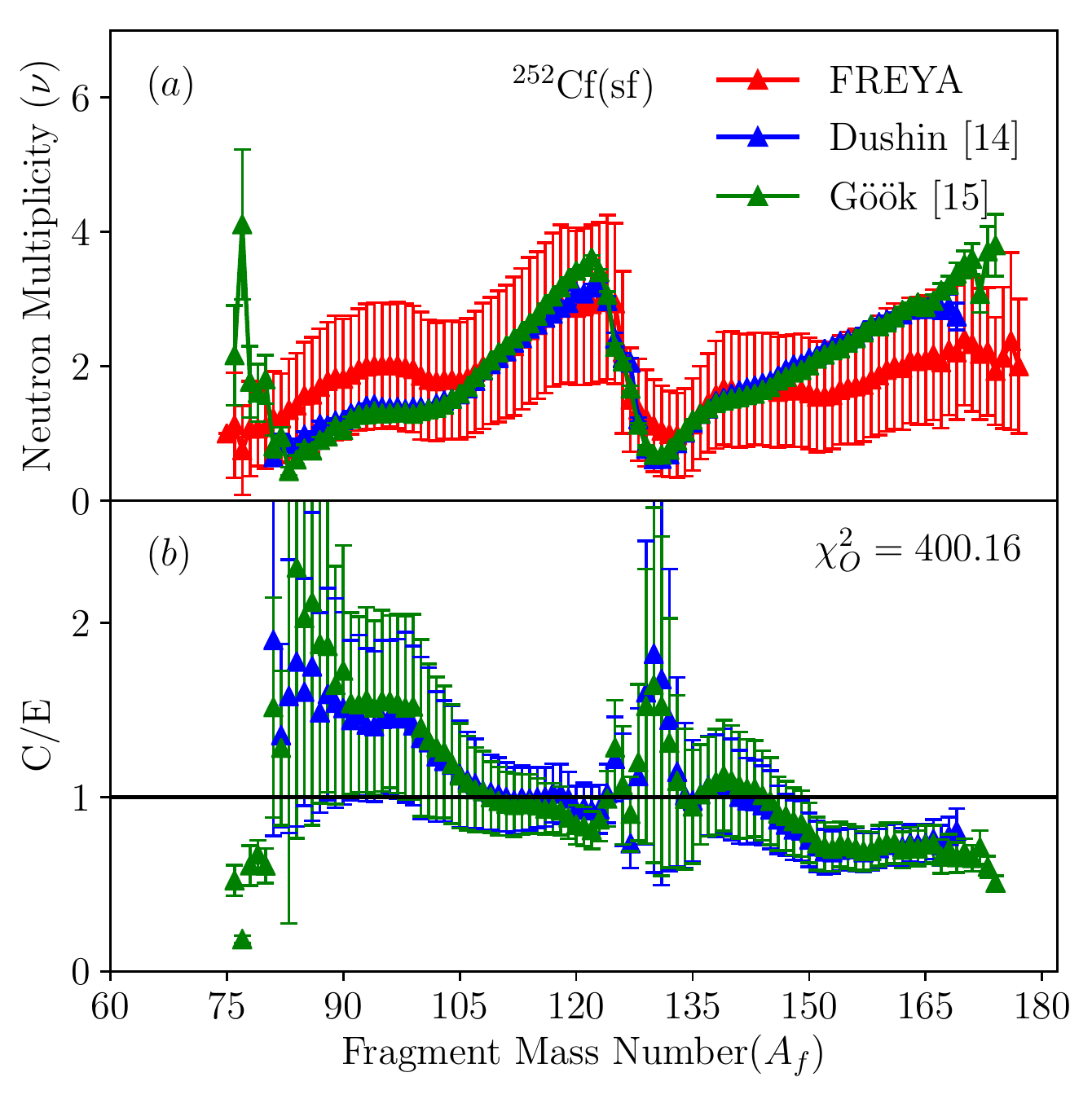}
\caption{(Color online) (a)
Neutron multiplicity as a function of fragment mass for \cf\ along with
experimental data from Refs.~\cite{Dushin,gook}.
As discussed in \Cref{sec:data}, we use the Dushin data in the
fit, and provide the G\"o\"ok data \cite{gook} for comparison.
Note that the uncertainty on the \code\ calculation is a calculation of the variance of the result,
and should therefore not be interpreted as the range of values we should expect \code\ to return.
(b) Ratio of calculated result from \code\ to the experimental results.}
\label{fig:n_Af}
\end{figure}

The \code\ results differ more significantly
from the data on the neutron multiplicity as a function of fragment mass, as seen in
\Cref{fig:n_Af}.
This is expected since $x$ is single valued and our fit employs the mass
region $105 < A < 145$.
Even though we have only used this well-behaved
region for our optimization procedure, it is important to note
that the result is still within the uncertainty on C/E, meaning that this
result is still statistically successful.
We can also compare to experimental data not used in the optimization.
In \Cref{fig:n_Af}, we show a more recent
data set for $\nu( A)$ which also agrees well with \code\ in the fit region.
Note that the uncertainty on the \code\ calculation is a calculation of the variance of the result,
and should therefore not be interpreted as the range of values we should expect \code\ to return.

\begin{figure}[t]
\centering
\includegraphics[width=0.49\textwidth]{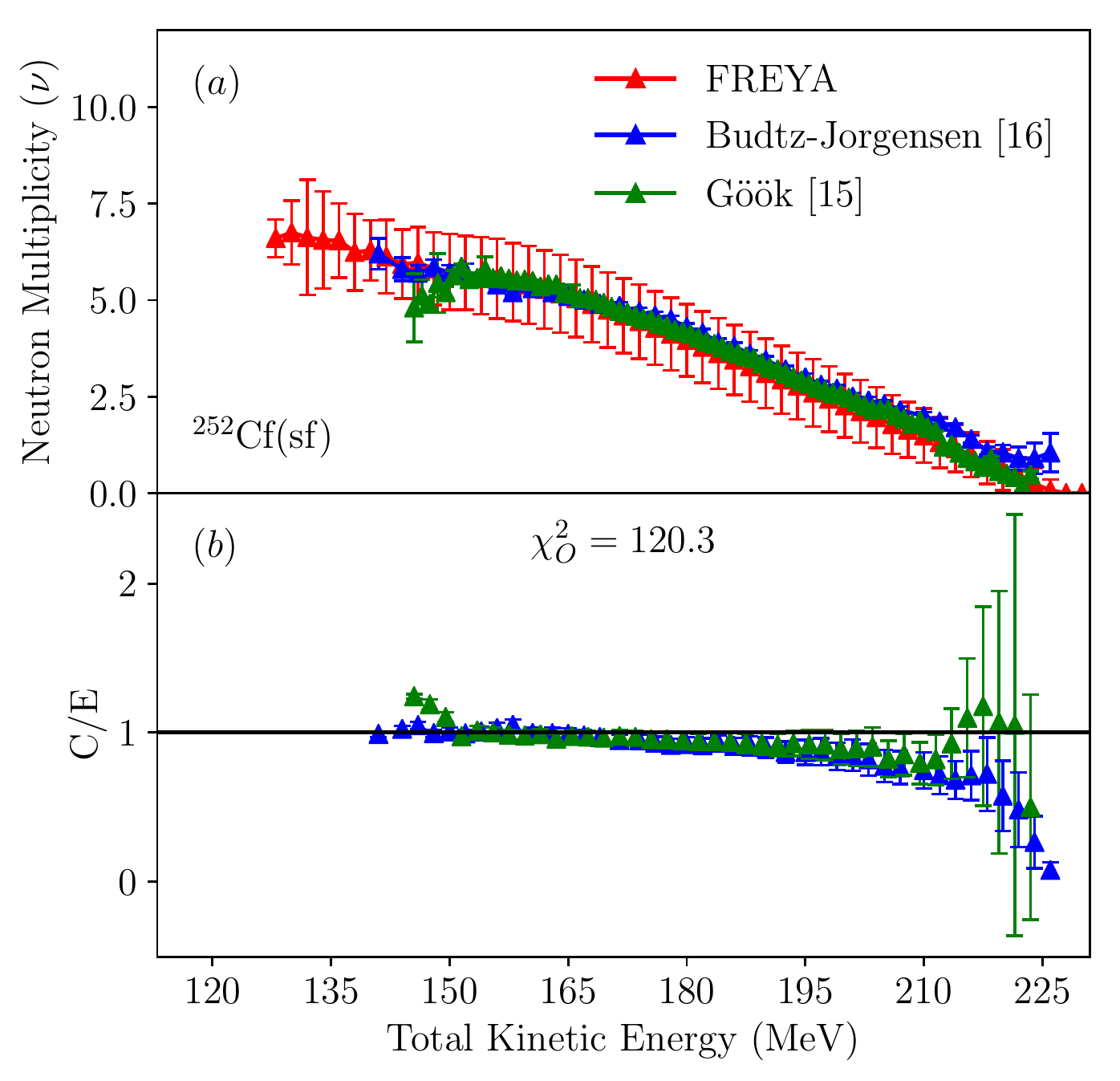}
\caption{(Color online) (a)
Neutron multiplicity as a function of total kinetic energy for \cf\
compared to experimental data from Refs.~\cite{gook,Bud}.
As is discussed in \Cref{sec:data}, we use the Budtz-Jorgensen data \cite{Bud}
for the fit, and provide the the G\"o\"ok data \cite{gook} for comparison.
Note that the uncertainty on the \code\ calculation is a calculation of the variance of the result,
and should therefore not be interpreted as the range of values we should expect \code\ to return.
(b) Ratio of calculated \code\ result to the experimental results.}
\label{fig:n_TKE}
\end{figure}

The results for the neutron multiplicity as a function of TKE in
\Cref{fig:n_TKE} are particularly successful for $160< \TKE < 190$ MeV.
In the region of low TKE, we see far fewer fragments,
so the results in this region are less reliable.
Similarly, as we move to higher TKE, while the results begin to
differ more, the uncertainty on C/E typically contains unity since there are also fewer events
with high TKE.
We also show the more recent data, not used in the fit, in
\Cref{fig:n_TKE}.  \code\ actually agrees better with this new data at large
\TKE\ because $\nu\left( \TKE \right) \rightarrow 0$ at large $\TKE$.

\begin{figure}[t]
\centering
\includegraphics[width=0.49\textwidth]{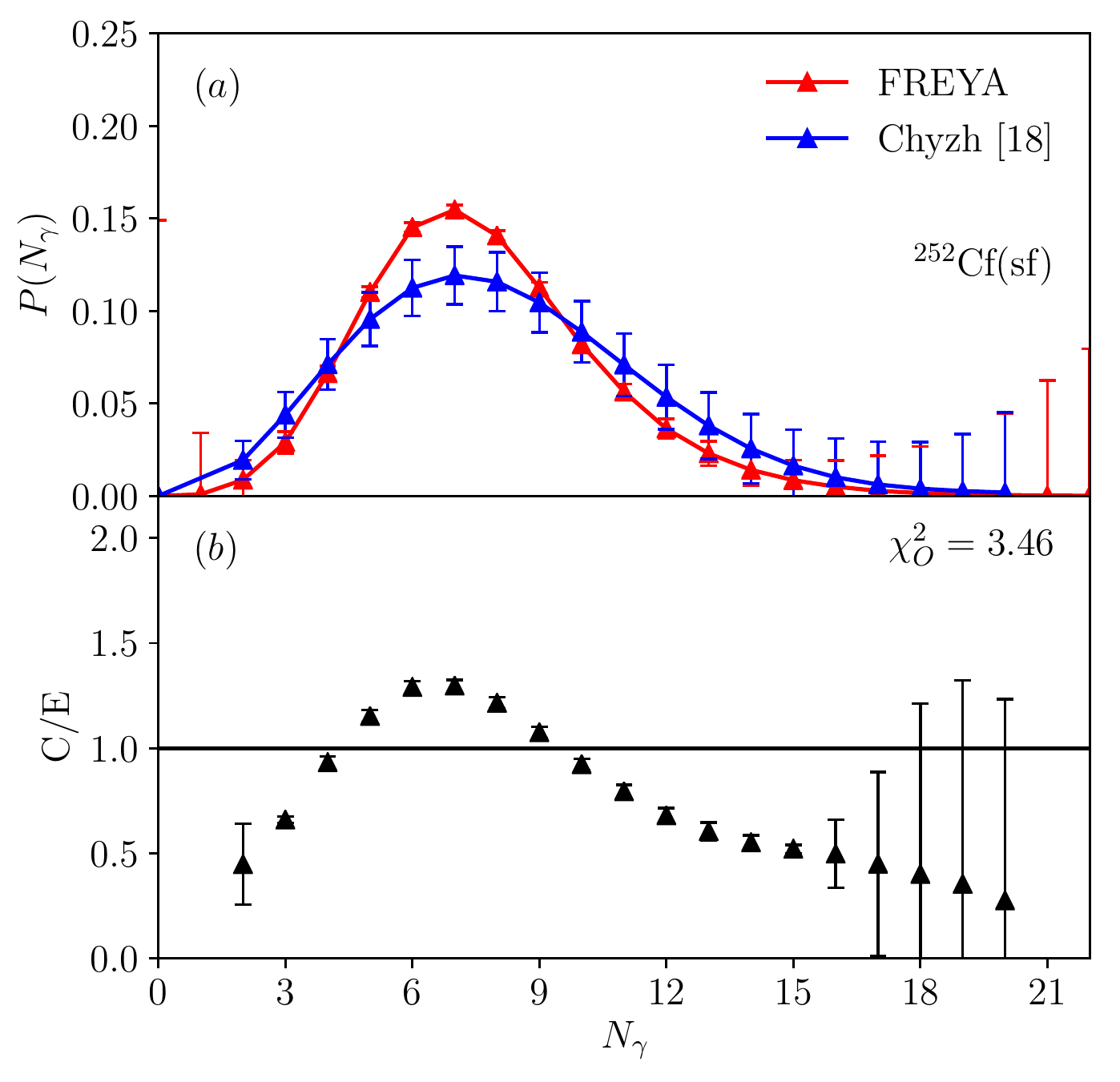}
\includegraphics[width=0.49\textwidth]{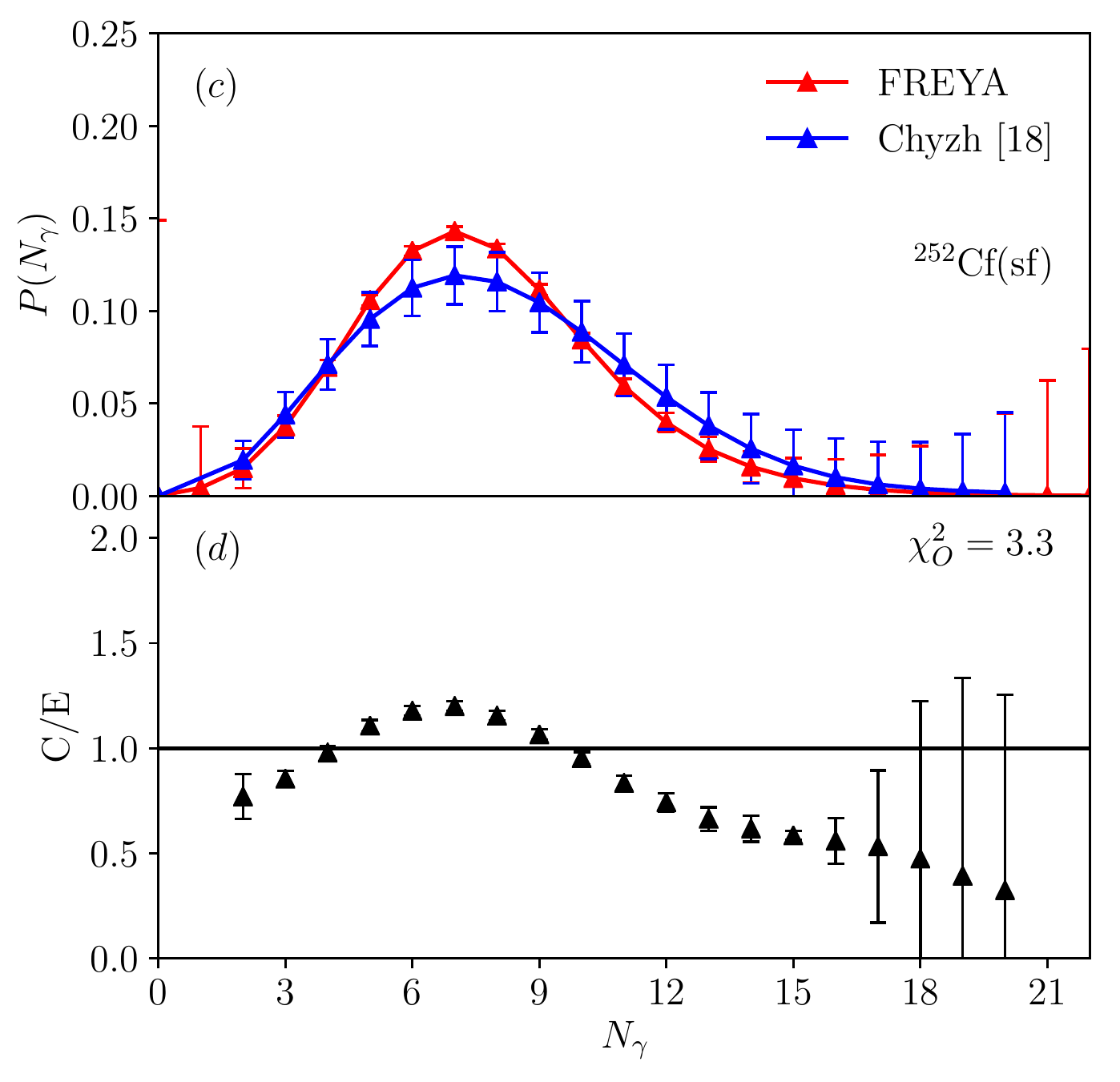}
\caption{(Color online) (a)
Gamma multiplicity distribution for \cf\ along with experimental data from \cite{Ch} before
correcting for multiple scattering.
(b) Ratio between the calculated values from \code\ and the experimental data.
(c) Gamma multiplicity distribution for \cf\ along with experimental data from
\cite{Ch} after
correcting for multiple scattering.
(d) Ratio between the calculated values from \code\ and the experimental data.}
\label{fig:m_mult}
\end{figure}

As explained in \Cref{sec:parameters}, the parameters, especially $c$, have a
high level of control over the shape of the neutron multiplicity distribution.
This is, however, not the case for the photon multiplicity distribution: there is
no parameter that has direct control over the width of this distribution as
there is for $P(\nu)$.
The shape generated by \code\ in \Cref{fig:m_mult} is narrower than the data.
There is an estimated uncertainty of $\pm 1$ in the
detected photon multiplicity \cite{Ch_comm}.
If we adjust the \code\ output to account for multiple scattering \cite{Ch_comm}, the width
becomes broader.
As we can see in \Cref{fig:m_mult}, after adjusting the \code\ output
for multiple scattering, the agreement of \code\ with the data is considerably
improved.  As is evident, the uncertainty on C/E is compatible with
unity in the high multiplicity range. The average photon multiplicity is also closely recreated.
These results can be found in \Cref{table:gammabar}, along with the average energy per photon.

\section{Conclusions}

We have performed a numerical optimization of the $5$
physics-based parameters in \code\ for all spontaneously-fissioning isotopes so far included.
The fits, using simulated annealing to find a global minimum,
which agree with our physics intuition, are also in rather good agreement with the empirical
values in \code\ 2.0.2 \cite{CPC_NVA}.

The parameters provide good agreement with the data where they are available.
We will next apply the fitting procedure we have developed here to neutron-induced fission.

\appendix

\section{$^{238,242}$Pu(sf) and $^{238}$U(sf) neutron multiplicity distributions}
\label{appendix:pofnu}

In this appendix, we show the neutron multiplicity distributions, $P(\nu)$
resulting from our fits to the \ur, \pu{238}, and \pu{242}\ evaluations by
Santi and Miller \cite{SM}.  The \cm\ multiplicity distribution is shown in the
next section, along with comparisons to other available \cm\ data.

\begin{figure}
\centering
\includegraphics[width=0.49\textwidth]{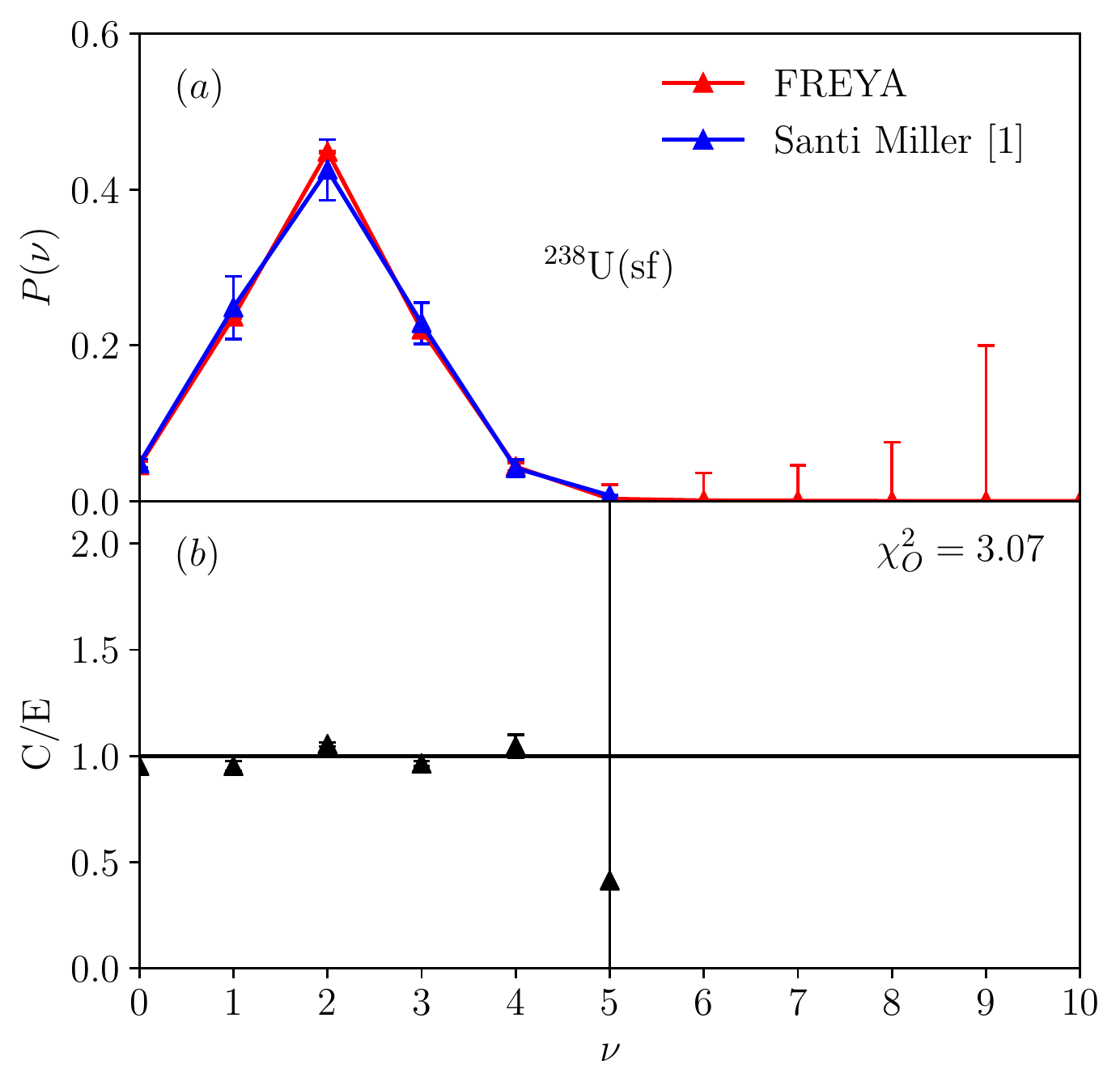}
\caption{(Color online) (a) Neutron multiplicity distribution for \ur\
is compared to the Santi-Miller evaluation Ref.~\cite{SM}.
Note that for the comparison, we use the square root of the uncertainty given in the evaluation,
since the actual reported uncertainty was so low it dominated the optimization.
(b) Ratio of calculated values to evaluation results.}
\label{fig:u238_p_nu}
\end{figure}

The three isotopes shown here, like \pu{240}, as shown in
Fig.~\ref{fig:pu240_p_nu},
are characterized by rather low average neutron multiplicities,
$\overline \nu \sim 2$ for
\ur\ and $\overline \nu \sim 2.15$ for $^{238,240,242}$Pu(sf).  These isotopes
are also distinguished by the lack of other data for optimization.  While
$^{240,242}$Pu(sf) have had recent measurements of the average photon
multiplicity and energy per photon, as shown in Table~\ref{table:gammabar},
the only data for optimization of the \code\ parameters for \ur\ and \pu{238}
are the Santi-Miller evaluations of $P(\nu)$ and the corresponding neutron
multiplicity moments.

\begin{figure}[t]
\centering
\includegraphics[width=0.49\textwidth]{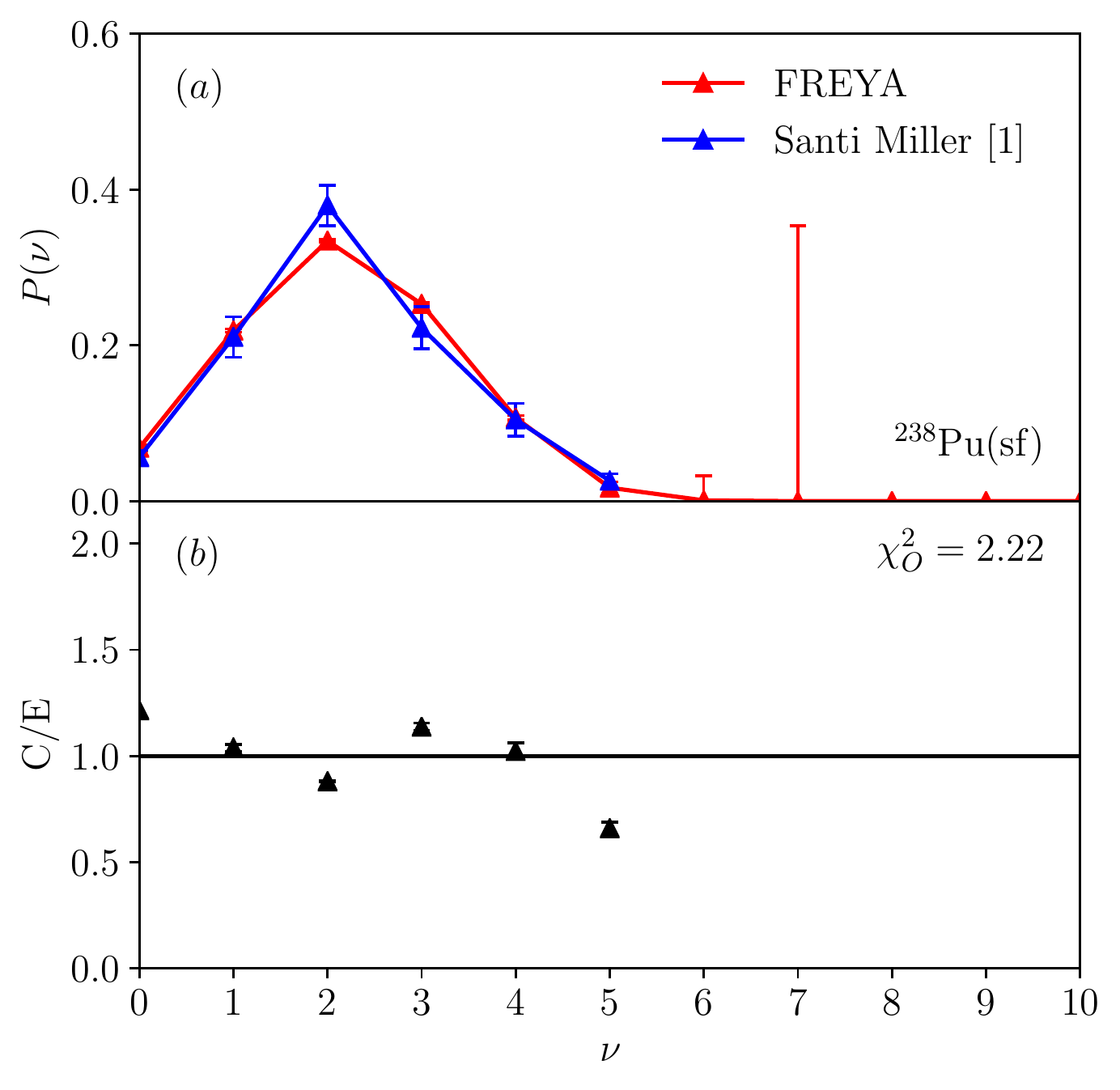}
\caption{(Color online) (a) Neutron multiplicity distribution for \pu{238}\
is compared to the Santi-Miller evaluation Ref.~\cite{SM}.
Note that for the comparison, we use the square root of the uncertainty given in the evaluation,
since the actual reported uncertainty was so low it dominated the optimization.
(b) Ratio of calculated values to evaluation results.}
\label{fig:pu238_p_nu}
\end{figure}

Figure~\ref{fig:u238_p_nu} shows the neutron multiplicity distribution
for \ur\ compared to the \code\ calculation.  This isotope, with the lowest
neutron multiplicity, is especially interesting because it is the only one with
$c < 1$.  Indeed, it is the only one where $\nu_3 < \nu_2$, with $\mu_2 = 2.87$
and $\nu_3 = 2.82$ respectively.  With the default value of $c = 1$, when the
neutron multiplicity is low, \code\ tends to produce $P(\nu)$ distributions
more narrowly peaked than the evaluations, requiring $c \sim 2-3$ for the
Pu(sf) isotopes included in \code, see Table~\ref{table:all_parameters}.
However, \ur\ is the only isotope where $c = 1$ produces a neutron multiplicity
distribution wider than the evaluation, requiring $c < 1$.  With the fitted
value of $c = 0.939$ in \code, there is good agreement with $P(\nu)$ as well
as with the moments of the distribution, see Table~\ref{table:nubar}.

\begin{figure}[t]
\centering
\includegraphics[width=0.49\textwidth]{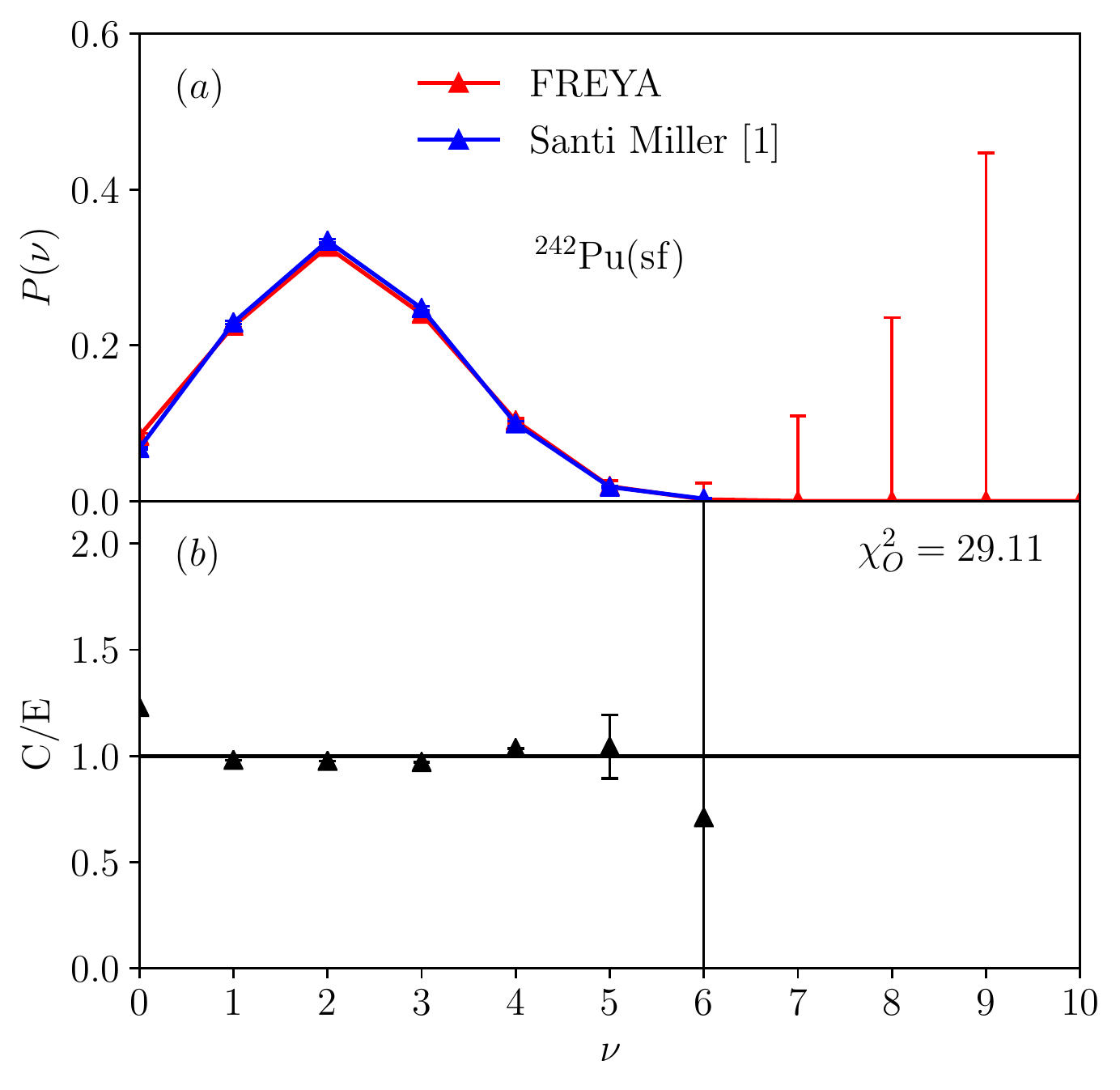}
\caption{(Color online) (a) Neutron multiplicity distribution for \pu{242}\
is compared to the Santi-Miller evaluation Ref.~\cite{SM}.
Note that for the comparison, we use the square root of the uncertainty given in the evaluation,
since the actual reported uncertainty was so low it dominated the optimization.
(b) Ratio of calculated values to evaluation results.}
\label{fig:pu242_p_nu}
\end{figure}

On the other hand, even though $\overline \nu$ for the Pu(sf) isotopes is only
$~7.5$\% larger than than of \ur, the higher moments are considerably larger,
leading to a broader $P(\nu)$.  For all three Pu(sf) isotopes, $\nu_2 \sim 3.8$
and $\nu_3 \sim 5.3$.  These multiplicity distributions, considerably broader
than a default $c = 1$ calculation in \code, result in the optimized values of
$c$ to be $\sim 2-3.4$, see Table~\ref{table:all_parameters}.  In each case,
$\nu_2$ is approximately 76\% larger than $\overline \nu$ while $\nu_3$
increases by $\sim 40$\% over $\nu_2$.

The evaluated multiplicity distributions are compared to the optimized \code\
results in Figs.~\ref{fig:pu238_p_nu} and \ref{fig:pu242_p_nu}.  The agreement
with \pu{242} and \code\ is better than that for \pu{238} where \code\
underestimates $\nu_3$ by 3\%.

\section{\cm\ results for $P(\nu)$, $\nu(A)$ and prompt fission neutron
  spectrum}

Because \cm\ has more data available for optimization than the evaluation of
$P(\nu)$ and the multiplicity moments, we have collected all the comparisons
of the \cm\ data with \code\ results in this appendix.  While there are data
on $\nu(A)$ and the prompt fission neutron spectrum for \cm, these data are not
of very high quality.  Nonetheless, they were useful for constraining the \code\
parameters and lead to results consistent with the other fits.

\begin{figure}[t]
\centering
\includegraphics[width=0.49\textwidth]{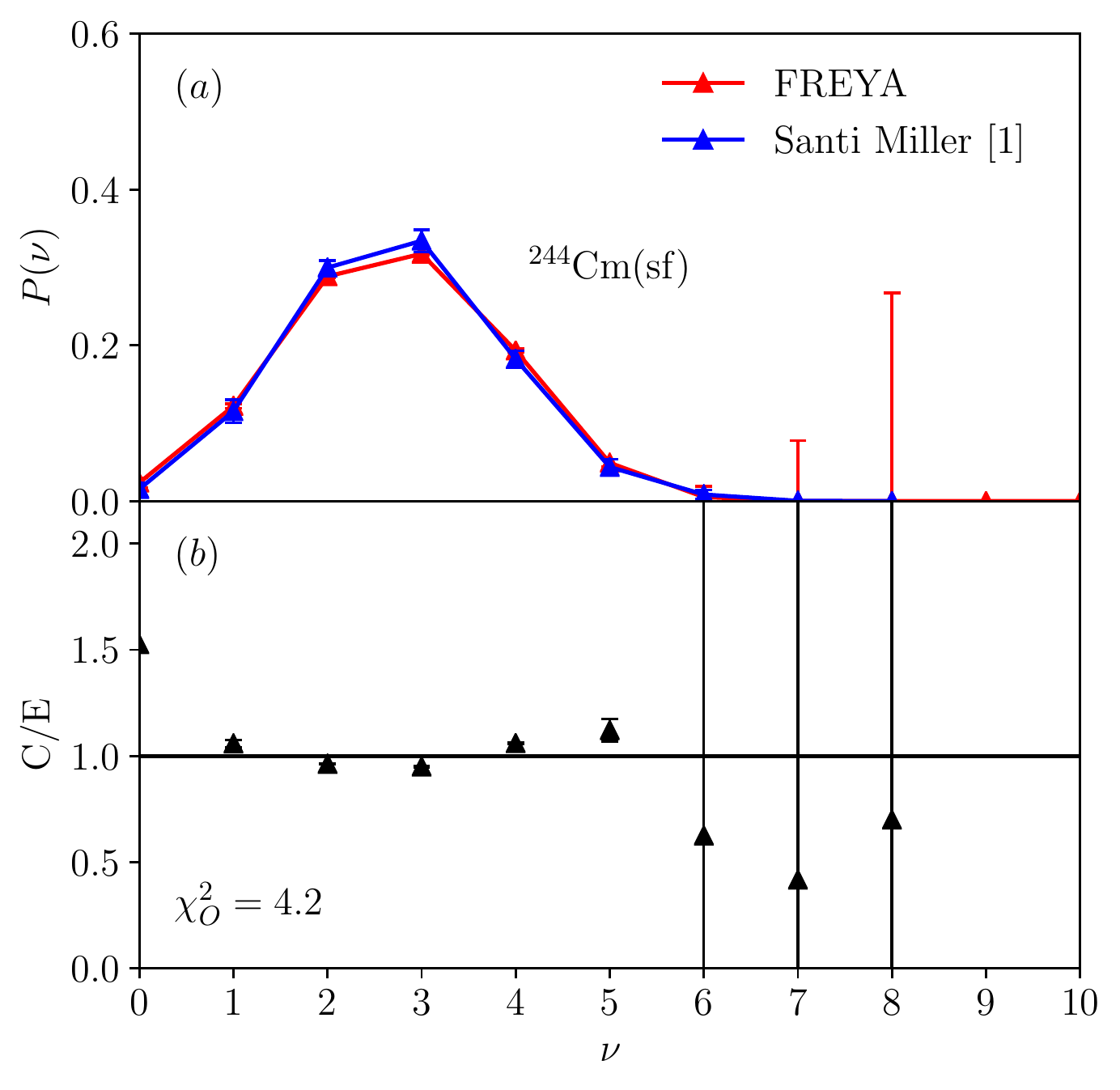}
\caption{(Color online) (a) Neutron multiplicity distribution for \cm\
is compared to the Santi-Miller evaluation Ref.~\cite{SM}.
Note that for the comparison, we use the square root of the uncertainty given in the evaluation,
since the actual reported uncertainty was so low it dominated the optimization.
(b) Ratio of calculated values to evaluation results.}
\label{fig:cm244_p_nu}
\end{figure}

We note that the \cm\ neutron multiplicity is considerably larger than those in
Appendix~\ref{appendix:pofnu}, $\overline \nu = 2.71$ relative to
$\overline \nu \sim 2-2.15$ for \ur\ and Pu(sf).  Consequently the behavior of
the moments of $P(\nu)$ are more similar to those of \cf:
$\nu_2/\overline \nu = 2.19$ for \cm\ and 3.18 for \cf\ while
$\nu_3/\nu_2 = 1.70$ for \cm\ and 2.65 for \cf.  In the case of the Pu isotopes,
$\nu_2/\overline \nu < 2$ and $\nu_3/\nu_2 \sim 1.4$, again emphasizing the
relative narrow multiplicity distributions attendant to smaller average neutron
multiplicities.  The optimal value for $c$ is thus reduced considerably for
\cm: $c = 1.391$, similar to the result for \cf\ of $c = 1.191$.  The comparison
with \code, shown in \Cref{fig:cm244_p_nu}, shows good agreement with the
evaluation of Ref.~\cite{SM}.

\begin{figure}[t]
\centering
\includegraphics[width=0.49\textwidth]{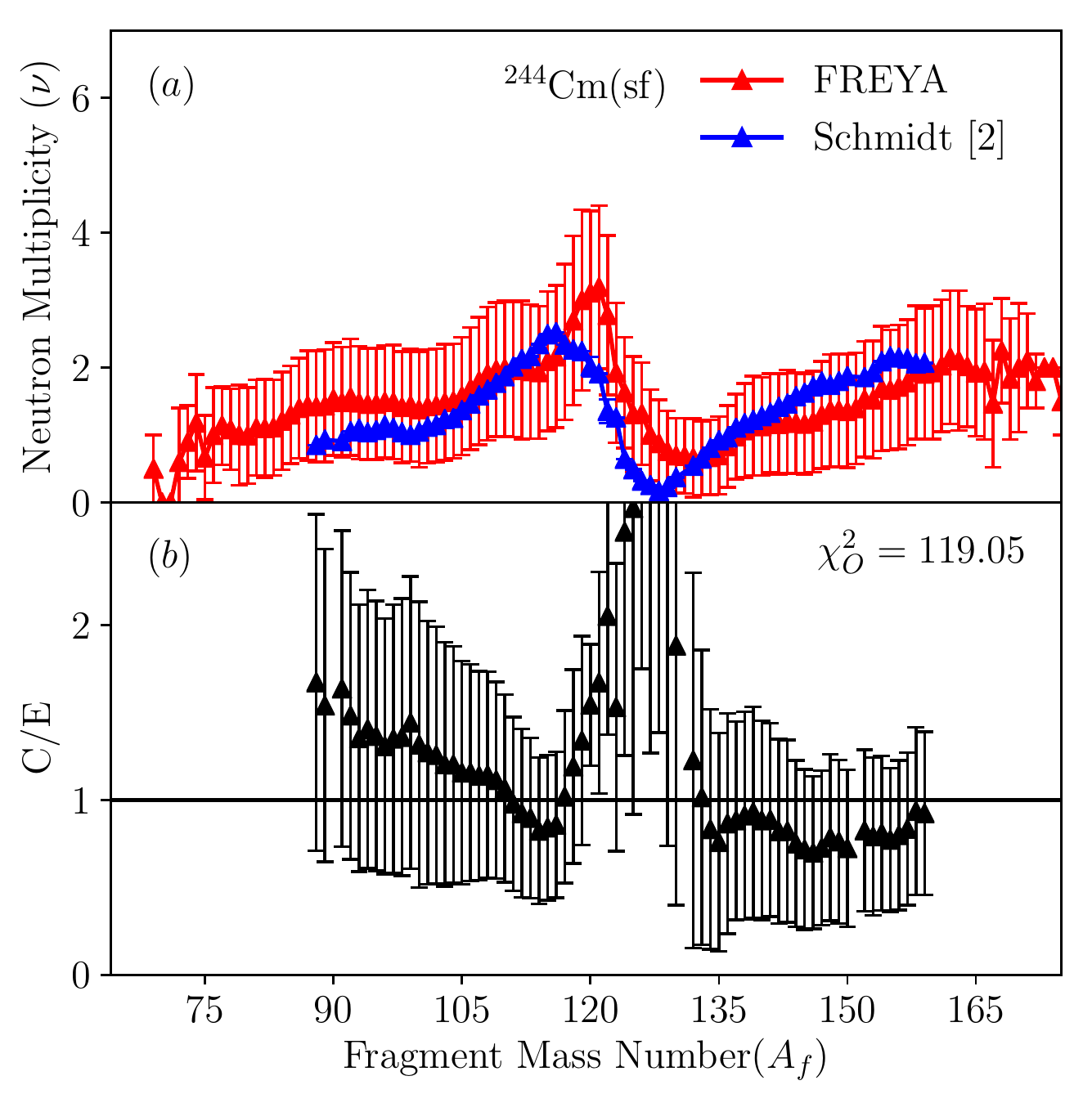}
\caption{(Color online) (a)
Neutron multiplicity as a function of fragment mass for \cm\ along with
experimental data from Refs.~\cite{schmidt}.
(b) Ratio of calculated result from \code\ to the experimental results.}
\label{fig:n_Af_cm}
\end{figure}

The neutron multiplicity as a function of fragment mass is shown compared to
\code\ in \Cref{fig:n_Af_cm}.  While the generic sawtooth pattern is recreated
well, there are some differences, especially near symmetry where the `tooth'
calculated with \code\ is sharper than that of the data.  The peak of $\nu(A)$
of the measured distribution is at a somewhat lighter mass number than in
\code.  Otherwise the agreement with the overall trends of the data away from
symmetry.  It is worth noting that Ref.~\cite{schmidt} made corrections to their
\cm\ data based on a \cf\ measurement taken with the same apparatus.  The
corrected $\nu(A)$, shown here, resulted in a significant backward shift of the 
light fragment peak near symmetry.

\begin{figure}[t]
\centering
\includegraphics[width=0.49\textwidth]{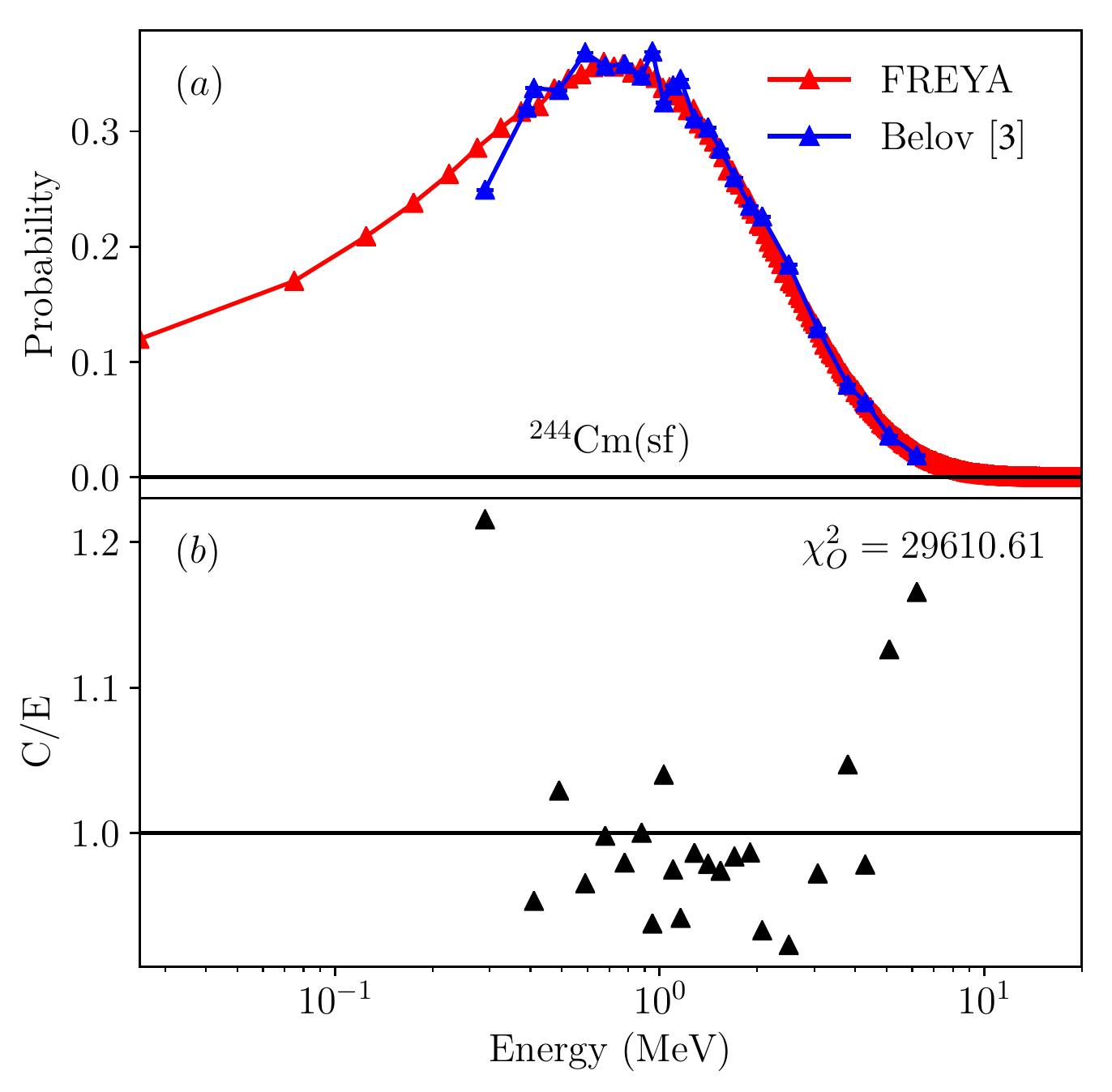}
\includegraphics[width=0.49\textwidth]{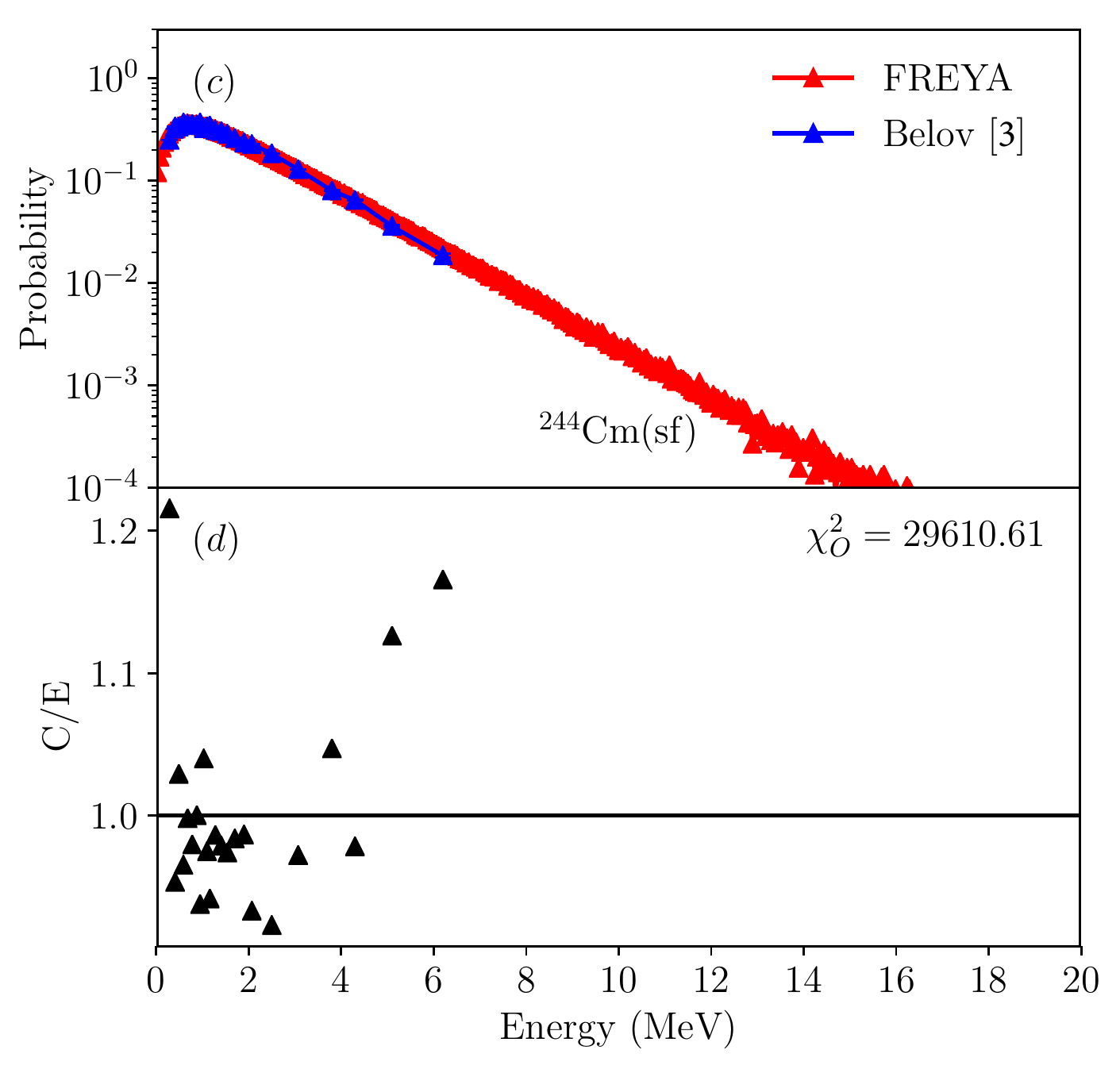}
\caption{(Color online) (a)
Neutron energy spectrum for \cm\ from \code\ as well as experimental data from \cite{belov}
with logarithmic scale on $x$ axis.
(b) Ratio of calculated result from \code\ to the experimental results.
In (c) and (d) the same results are shown
now with a logarithmic scale on the $y$-axis, and a linear scale on the $x$-axis.}
\label{fig:cm_n_spec}
\end{figure}

Finally, we compare \code\ to a measurement of the \cm\ prompt fission
neutron energy spectrum in \Cref{fig:cm_n_spec}.
The measured energy range is rather narrow, with a good deal of scatter between
the points and a drop off of the lowest energy point.  Nonetheless the agreement
of the data with the calculation is rather good over the common energy interval.

\section*{Acknowledgments}

We wish to acknowledge helpful conversations with Andrew Nicholson,
Jorgen Randrup, Jerome Verbeke, and Patrick Talou.
The computational work was done with the Savio cluster using the faculty computing allowance provided by
Berkeley Research Computing.
This work was supported by the Office of Nuclear Physics
in the U.S. Department of Energy's office of Science under contracts
No. DE-AC52-07NA27344 (RV) and
DE-AC02-05CH11231 (LAB),
as well as by the U.S. Department of Energy's National Nuclear
Security Administration.

 % do not change
\end{document}